\documentclass[11pt]{article}

\usepackage[utf8]{inputenc}
\usepackage[T1]{fontenc}
\usepackage{lmodern}
\usepackage[letterpaper,margin=1in]{geometry}
\usepackage{microtype}
\usepackage{graphicx}
\usepackage{booktabs}
\usepackage{amsmath,amssymb}
\usepackage{array}
\usepackage{longtable}
\usepackage{float}
\usepackage{placeins}
\usepackage{authblk}
\usepackage{tikz}
\usepackage[numbers,sort&compress]{natbib}
\usepackage{xurl}
\usepackage{hyperref}
\usepackage{listings}
\usepackage[most]{tcolorbox}

\definecolor{ToubaCodeAccent}{HTML}{2457C5}
\definecolor{ToubaCodeBorder}{HTML}{C9D7F2}
\definecolor{ToubaCodeBackground}{HTML}{F6F8FC}
\definecolor{ToubaCodeHeader}{HTML}{EDF3FF}
\definecolor{ToubaCodeString}{HTML}{176B65}

\lstdefinestyle{ToubaListing}{
  basicstyle=\ttfamily\small\color{black!86},
  columns=fullflexible,
  keepspaces=true,
  showstringspaces=false,
  breaklines=true,
  breakatwhitespace=true,
  tabsize=2,
  keywordstyle=\color{ToubaCodeAccent}\bfseries,
  stringstyle=\color{ToubaCodeString},
  morekeywords={import,from,export,const}
}
\lstdefinelanguage{Touba}{
  sensitive=true,
  morekeywords={module,tool,agent,strategy,description,takes,as,gives,
    responsibility,uses,callable_by,do,with,into,give,memory,write,requires,secrets},
  morestring=[b]"
}
\lstdefinestyle{ToubaSource}{
  style=ToubaListing,
  language=Touba,
  basicstyle=\ttfamily\small\color{black!86},
  numbers=left,
  numberstyle=\scriptsize\color{black!45},
  numbersep=9pt,
  xleftmargin=12pt
}

\newtcblisting{ToubaCode}[1][]{
  enhanced,
  breakable,
  listing only,
  colback=ToubaCodeBackground,
  colframe=ToubaCodeBorder,
  colbacktitle=ToubaCodeHeader,
  coltitle=ToubaCodeAccent,
  fonttitle=\sffamily\small\bfseries,
  boxrule=0.55pt,
  arc=4pt,
  left=9pt,
  right=9pt,
  top=7pt,
  bottom=7pt,
  toptitle=5pt,
  bottomtitle=5pt,
  titlerule=0pt,
  borderline west={2.2pt}{0pt}{ToubaCodeAccent},
  before skip=10pt,
  after skip=12pt,
  listing options={style=ToubaListing},
  #1
}

\usetikzlibrary{arrows.meta,positioning,fit}

\newcommand{\PaperTitle}{Responsibility Boundaries for Agentic Systems: A Comparative Evaluation of Probabilistic Agentic Programming}
\newcommand{\PaperTitleBreak}{Responsibility Boundaries for Agentic Systems:\\A Comparative Evaluation of Probabilistic Agentic Programming}
\newcommand{\PaperPdfAuthor}{Khadim Lo, Soukaina Alaoui Ismaili, and Jean-Sebastien Dessureault}
\newcommand{\PaperKeywords}{Probabilistic Agentic Programming; probabilistic agentic systems; agent-oriented programming; governed tool use; capability boundaries; structured execution traces; auditability}

\newcommand{\PaperAuthorContributions}{K.L. developed the comparative argument for PAP, the comparison dimensions, and the Touba-aligned analysis. S.A.I. contributed to research discussion, manuscript development, and critical review. J.-S.D. provided academic and research supervision, theoretical grounding, and critical review.}

\newcommand{\PaperCompetingInterestsDeclaration}{The authors are members of the research team developing Touba, the implementation evaluated in this study. They designed the evaluation and analyzed its results.}

\newcommand{\PaperAIUseDeclaration}{AI-assisted tools were used to improve the wording and clarity of the manuscript.}


\hypersetup{
  colorlinks=true,
  linkcolor=black,
  citecolor=black,
  urlcolor=blue,
  pdftitle={\PaperTitle},
  pdfauthor={\PaperPdfAuthor}
}

\title{\PaperTitleBreak\thanks{Touba documentation: \url{https://toubalang.org/docs/}.}}
\author[1]{Khadim Lo}
\author[1]{Soukaina Alaoui Ismaili}
\author[2]{Jean-S\'{e}bastien Dessureault}
\affil[1]{Electrical \& Computer Engineering, LSSI, UQTR\\
\texttt{khadim.lo@uqtr.ca}; \texttt{soukaina.alaoui.ismaili@uqtr.ca}}
\affil[2]{Mathematics \& Computer Science, LSSI, UQTR\\
\texttt{jean-sebastien.dessureault@uqtr.ca}}
\date{October 2026}

\newcommand{\PaperPrintKeywords}{\noindent\textbf{Keywords:} \PaperKeywords.}

\newcommand{\PaperPrintDeclarations}{%
  \section*{Author Contributions}
  \PaperAuthorContributions
  \section*{Author Relationship to the Evaluated System}
  \PaperCompetingInterestsDeclaration
  \section*{AI Assistance}
  \PaperAIUseDeclaration
}
\newcommand{\PaperPrintBibliography}{%
  \begin{thebibliography}{61}
\providecommand{\natexlab}[1]{#1}
\providecommand{\url}[1]{\texttt{#1}}
\expandafter\ifx\csname urlstyle\endcsname\relax
  \providecommand{\doi}[1]{doi: #1}\else
  \providecommand{\doi}{doi: \begingroup \urlstyle{rm}\Url}\fi

\bibitem[{Anthropic}(2026)]{anthropicToolUse2026}
{Anthropic}.
\newblock Tool use with {Claude}, 2026.
\newblock URL
  \url{https://platform.claude.com/docs/en/agents-and-tools/tool-use/overview}.

\bibitem[Backus(1978)]{backus1978vonNeumann}
John Backus.
\newblock Can programming be liberated from the {von Neumann} style? a
  functional style and its algebra of programs.
\newblock \emph{Communications of the ACM}, 21\penalty0 (8):\penalty0 613--641,
  1978.
\newblock \doi{10.1145/359576.359579}.

\bibitem[Black(2013)]{black2013oopHistory}
Andrew~P. Black.
\newblock Object-oriented programming: Some history, and challenges for the
  next fifty years.
\newblock \emph{Information and Computation}, 231:\penalty0 3--20, 2013.
\newblock \doi{10.1016/j.ic.2013.08.002}.

\bibitem[Cutler et~al.(2024)Cutler, Disselkoen, Eline, He, Headley, Hicks,
  Hietala, Ioannidis, Kastner, Mamat, McAdams, McCutchen, Rungta, Torlak, and
  Wells]{cedar2024authorizationLanguage}
Joseph~W. Cutler, Craig Disselkoen, Aaron Eline, Shaobo He, Kyle Headley,
  Michael Hicks, Kesha Hietala, Eleftherios Ioannidis, John Kastner, Anwar
  Mamat, Darin McAdams, Matt McCutchen, Neha Rungta, Emina Torlak, and
  Andrew~M. Wells.
\newblock {Cedar}: A new language for expressive, fast, safe, and analyzable
  authorization.
\newblock \emph{Proceedings of the ACM on Programming Languages}, 8\penalty0
  (OOPSLA1):\penalty0 118:1--118:28, 2024.
\newblock \doi{10.1145/3649835}.
\newblock URL
  \url{https://cdn.amazon.science/96/a8/1b427993481cbdf0ef2c8ca6db85/cedar-a-new-language-for-expressive-fast-safe-and-analyzable-authorization.pdf}.

\bibitem[Dastani(2008)]{dastani2008twoAPL}
Mehdi Dastani.
\newblock {2APL}: A practical agent programming language.
\newblock \emph{Autonomous Agents and Multi-Agent Systems}, 16\penalty0
  (3):\penalty0 214--248, 2008.
\newblock \doi{10.1007/s10458-008-9036-y}.
\newblock URL
  \url{https://link.springer.com/content/pdf/10.1007/s10458-008-9036-y.pdf}.

\bibitem[Debenedetti et~al.(2024)Debenedetti, Zhang, Balunovi{\'c},
  Beurer-Kellner, Fischer, and Tram{\`e}r]{agentdojo2024benchmark}
Edoardo Debenedetti, Jie Zhang, Mislav Balunovi{\'c}, Luca Beurer-Kellner, Marc
  Fischer, and Florian Tram{\`e}r.
\newblock {AgentDojo}: A dynamic environment to evaluate prompt injection
  attacks and defenses for {LLM} agents.
\newblock In \emph{Advances in Neural Information Processing Systems},
  volume~37, 2024.
\newblock \doi{10.52202/079017-2636}.
\newblock URL
  \url{https://proceedings.neurips.cc/paper_files/paper/2024/file/97091a5177d8dc64b1da8bf3e1f6fb54-Paper-Datasets_and_Benchmarks_Track.pdf}.
\newblock Datasets and Benchmarks Track.

\bibitem[Debenedetti et~al.(2025)Debenedetti, Shumailov, Fan, Hayes, Carlini,
  Fabian, Kern, Shi, Terzis, and
  Tram{\`e}r]{camel2025defeatingPromptInjections}
Edoardo Debenedetti, Ilia Shumailov, Tianqi Fan, Jamie Hayes, Nicholas Carlini,
  Daniel Fabian, Christoph Kern, Chongyang Shi, Andreas Terzis, and Florian
  Tram{\`e}r.
\newblock Defeating prompt injections by design.
\newblock \emph{arXiv}, 2025.
\newblock \doi{10.48550/arXiv.2503.18813}.
\newblock CaMeL; arXiv v2, 24 June 2025.

\bibitem[Dijkstra(1968)]{dijkstra1968goto}
Edsger~W. Dijkstra.
\newblock Go to statement considered harmful.
\newblock \emph{Communications of the ACM}, 11\penalty0 (3):\penalty0 147--148,
  1968.
\newblock \doi{10.1145/362929.362947}.

\bibitem[{European Parliament and Council of the European
  Union}(2024)]{eu2024aiAct}
{European Parliament and Council of the European Union}.
\newblock Regulation ({EU}) 2024/1689 laying down harmonised rules on
  artificial intelligence, 2024.
\newblock URL \url{https://eur-lex.europa.eu/eli/reg/2024/1689/oj}.
\newblock Official Journal of the European Union, 12 July 2024.

\bibitem[{Executive Office of the President, Office of Management and
  Budget}(2024)]{omb2024m2410}
{Executive Office of the President, Office of Management and Budget}.
\newblock Advancing governance, innovation, and risk management for agency use
  of artificial intelligence, 2024.
\newblock URL
  \url{https://bidenwhitehouse.archives.gov/wp-content/uploads/2024/03/M-24-10-Advancing-Governance-Innovation-and-Risk-Management-for-Agency-Use-of-Artificial-Intelligence.pdf}.
\newblock Memorandum M-24-10, 28 March 2024.

\bibitem[{Executive Office of the President, Office of Management and
  Budget}(2025)]{omb2025m2521}
{Executive Office of the President, Office of Management and Budget}.
\newblock Accelerating federal use of {AI} through innovation, governance, and
  public trust, 2025.
\newblock URL
  \url{https://www.whitehouse.gov/wp-content/uploads/2025/02/M-25-21-Accelerating-Federal-Use-of-AI-through-Innovation-Governance-and-Public-Trust.pdf}.
\newblock Memorandum M-25-21, 3 April 2025; rescinds and replaces M-24-10.

\bibitem[{Google}(2026{\natexlab{a}})]{googleAdkDocs2026}
{Google}.
\newblock Agent development kit documentation, 2026{\natexlab{a}}.
\newblock URL \url{https://adk.dev/}.

\bibitem[{Google}(2026{\natexlab{b}})]{googleGeminiFunctionCalling2026}
{Google}.
\newblock Function calling with the {Gemini API}, 2026{\natexlab{b}}.
\newblock URL \url{https://ai.google.dev/gemini-api/docs/function-calling}.

\bibitem[Hardy(1988)]{hardy1988confusedDeputy}
Norm Hardy.
\newblock The confused deputy (or why capabilities might have been invented).
\newblock \emph{ACM SIGOPS Operating Systems Review}, 22\penalty0 (4):\penalty0
  36--38, 1988.
\newblock \doi{10.1145/54289.871709}.
\newblock URL
  \url{https://www.cs.utexas.edu/~witchel/S25-380L/papers/hardy88confused.pdf}.

\bibitem[Hewitt et~al.(1973)Hewitt, Bishop, and Steiger]{hewitt1973actors}
Carl Hewitt, Peter Bishop, and Richard Steiger.
\newblock A universal modular actor formalism for artificial intelligence.
\newblock In \emph{Proceedings of the 3rd International Joint Conference on
  Artificial Intelligence}, pages 235--245, 1973.
\newblock URL \url{https://www.ijcai.org/Proceedings/73/Papers/027B.pdf}.
\newblock IJCAI-73 proceedings.

\bibitem[Hofer et~al.(2026)Hofer, Debenedetti, and
  Tram{\`e}r]{hofer2026automatedPromptInjection}
David Hofer, Edoardo Debenedetti, and Florian Tram{\`e}r.
\newblock Assessing automated prompt injection attacks in agentic environments.
\newblock \emph{arXiv}, 2026.
\newblock \doi{10.48550/arXiv.2606.10525}.

\bibitem[{in-toto}(2026)]{inToto2026Framework}
{in-toto}.
\newblock A framework to secure the integrity of software supply chains, 2026.
\newblock URL \url{https://in-toto.io/}.

\bibitem[Karpas et~al.(2022)Karpas, Abend, Belinkov, Lenz, Lieber, Ratner,
  Shoham, Bata, Levine, Leyton-Brown, Muhlgay, Rozen, Schwartz, Shachaf,
  Shalev-Shwartz, Shashua, and Tenenholtz]{karpas2022mrklSystems}
Ehud Karpas, Omri Abend, Yonatan Belinkov, Barak Lenz, Opher Lieber, Nir
  Ratner, Yoav Shoham, Hofit Bata, Yoav Levine, Kevin Leyton-Brown, Dor
  Muhlgay, Noam Rozen, Erez Schwartz, Gal Shachaf, Shai Shalev-Shwartz, Amnon
  Shashua, and Moshe Tenenholtz.
\newblock {MRKL} systems: A modular, neuro-symbolic architecture that combines
  large language models, external knowledge sources and discrete reasoning.
\newblock \emph{arXiv}, 2022.
\newblock \doi{10.48550/arXiv.2205.00445}.

\bibitem[{LangChain}(2026{\natexlab{a}})]{langchainJsReview2026}
{LangChain}.
\newblock {LangChain JavaScript}: Human-in-the-loop, 2026{\natexlab{a}}.
\newblock URL
  \url{https://docs.langchain.com/oss/javascript/langchain/human-in-the-loop}.

\bibitem[{LangChain}(2026{\natexlab{b}})]{langgraphJsInterrupts2026}
{LangChain}.
\newblock {LangGraph JavaScript}: Interrupts, 2026{\natexlab{b}}.
\newblock URL
  \url{https://docs.langchain.com/oss/javascript/langgraph/interrupts}.

\bibitem[{LangChain}(2026{\natexlab{c}})]{langgraphPersistence2026}
{LangChain}.
\newblock {LangGraph} persistence, 2026{\natexlab{c}}.
\newblock URL
  \url{https://docs.langchain.com/oss/python/langgraph/persistence}.

\bibitem[Li et~al.(2026)Li, Wen, Shi, Zhang, Vorobeychik, and
  Xiao]{agentdyn2026benchmark}
Hao Li, Ruoyao Wen, Shanghao Shi, Ning Zhang, Yevgeniy Vorobeychik, and Chaowei
  Xiao.
\newblock {AgentDyn}: Are your agent security defenses deployable in real-world
  dynamic environments?
\newblock \emph{arXiv}, 2026.
\newblock \doi{10.48550/arXiv.2602.03117}.
\newblock URL \url{https://arxiv.org/abs/2602.03117v3}.
\newblock arXiv v3, 7 May 2026.

\bibitem[Lo et~al.(2026)Lo, Alaoui~Ismaili, and
  Dessureault]{lo2026foundationalPAP}
Khadim Lo, Soukaina Alaoui~Ismaili, and Jean-S{\'e}bastien Dessureault.
\newblock Probabilistic agentic programming: A foundational paradigm for
  governed autonomy.
\newblock Manuscript submitted to AI and Ethics, 2026.

\bibitem[Lucassen and Gifford(1988)]{lucassen1988effects}
John~M. Lucassen and David~K. Gifford.
\newblock Polymorphic effect systems.
\newblock In \emph{Proceedings of the 15th ACM SIGPLAN-SIGACT Symposium on
  Principles of Programming Languages}, pages 47--57, 1988.
\newblock \doi{10.1145/73560.73564}.
\newblock URL \url{https://cs.ioc.ee/ewscs/2010/mycroft/lucassen-popl88.pdf}.

\bibitem[Lynch and Vaandrager(1995)]{lynch1995simulations}
Nancy Lynch and Frits Vaandrager.
\newblock Forward and backward simulations: {I}. untimed systems.
\newblock \emph{Information and Computation}, 121\penalty0 (2):\penalty0
  214--233, 1995.
\newblock \doi{10.1006/inco.1995.1134}.
\newblock URL \url{https://ir.cwi.nl/pub/1393/1393D.pdf}.

\bibitem[Miller(2006)]{miller2006robustComposition}
Mark~Samuel Miller.
\newblock \emph{Robust Composition: Towards a Unified Approach to Access
  Control and Concurrency Control}.
\newblock PhD thesis, Johns Hopkins University, 2006.
\newblock URL \url{https://jscholarship.library.jhu.edu/handle/1774.2/873}.

\bibitem[{Model Context Protocol}(2025)]{anthropicModelContextProtocol2025}
{Model Context Protocol}.
\newblock Model context protocol specification, 2025.
\newblock URL \url{https://modelcontextprotocol.io/specification/2025-11-25}.
\newblock Version 2025-11-25.

\bibitem[{National Institute of Standards and Technology}(2023)]{nist2023aiRmf}
{National Institute of Standards and Technology}.
\newblock Artificial intelligence risk management framework ({AI RMF} 1.0).
\newblock {NIST AI} 100-1, National Institute of Standards and Technology,
  2023.
\newblock URL \url{https://doi.org/10.6028/NIST.AI.100-1}.
\newblock Published 26 January 2023.

\bibitem[{National Institute of Standards and
  Technology}(2024)]{nist2024genAiProfile}
{National Institute of Standards and Technology}.
\newblock Artificial intelligence risk management framework: Generative
  artificial intelligence profile.
\newblock {NIST AI} 600-1, National Institute of Standards and Technology,
  2024.
\newblock URL \url{https://doi.org/10.6028/NIST.AI.600-1}.
\newblock Published 26 July 2024.

\bibitem[Necula(1997)]{necula1997proofCarryingCode}
George~C. Necula.
\newblock Proof-carrying code.
\newblock In \emph{Proceedings of the 24th ACM SIGPLAN-SIGACT Symposium on
  Principles of Programming Languages}, pages 106--119, 1997.
\newblock \doi{10.1145/263699.263712}.
\newblock URL
  \url{https://homes.cs.washington.edu/%7Emernst/teaching/6.893/readings/necula-popl97.pdf}.

\bibitem[{OASIS eXtensible Access Control Markup Language Technical
  Committee}(2013)]{oasis2013xacml}
{OASIS eXtensible Access Control Markup Language Technical Committee}.
\newblock {eXtensible Access Control Markup Language} ({XACML}) version 3.0,
  2013.
\newblock URL
  \url{https://docs.oasis-open.org/xacml/3.0/xacml-3.0-core-spec-os-en.html}.
\newblock OASIS Standard, 22 January 2013.

\bibitem[{OASIS Web Services Business Process Execution Language Technical
  Committee}(2007)]{oasis2007wsbpel}
{OASIS Web Services Business Process Execution Language Technical Committee}.
\newblock Web services business process execution language version 2.0, 2007.
\newblock URL
  \url{https://docs.oasis-open.org/wsbpel/2.0/OS/wsbpel-v2.0-OS.html}.
\newblock OASIS Standard, 11 April 2007.

\bibitem[{Office of the Superintendent of Financial
  Institutions}(2025)]{osfi2025e23}
{Office of the Superintendent of Financial Institutions}.
\newblock Guideline {E-23}: Model risk management, 2025.
\newblock URL
  \url{https://www.osfi-bsif.gc.ca/en/guidance/guidance-library/guideline-e-23-model-risk-management-2027}.
\newblock Published 11 September 2025; effective 1 May 2027.

\bibitem[{Open Policy Agent}(2026)]{opa2026docs}
{Open Policy Agent}.
\newblock Open policy agent documentation, 2026.
\newblock URL \url{https://www.openpolicyagent.org/docs}.

\bibitem[{OpenAI}(2026{\natexlab{a}})]{openaiAgentsSdk2026}
{OpenAI}.
\newblock {OpenAI Agents SDK} documentation, 2026{\natexlab{a}}.
\newblock URL \url{https://developers.openai.com/api/docs/guides/agents}.

\bibitem[{OpenAI}(2026{\natexlab{b}})]{openaiSdkReview2026}
{OpenAI}.
\newblock Agents {SDK}: Guardrails and human review, 2026{\natexlab{b}}.
\newblock URL
  \url{https://developers.openai.com/api/docs/guides/agents/guardrails-approvals}.

\bibitem[{OpenAI}(2026{\natexlab{c}})]{openaiSdkTracing2026}
{OpenAI}.
\newblock Agents {SDK}: Integrations and observability, 2026{\natexlab{c}}.
\newblock URL
  \url{https://developers.openai.com/api/docs/guides/agents/integrations-observability}.

\bibitem[{OpenTelemetry}(2026)]{opentelemetryGenAi2026}
{OpenTelemetry}.
\newblock Semantic conventions for generative {AI} systems, 2026.
\newblock URL
  \url{https://github.com/open-telemetry/semantic-conventions-genai/tree/main/docs}.

\bibitem[{OWASP GenAI Security
  Project}(2025{\natexlab{a}})]{owasp2025agenticTop10}
{OWASP GenAI Security Project}.
\newblock {OWASP GenAI} security project releases top 10 risks and mitigations
  for agentic {AI} security, 2025{\natexlab{a}}.
\newblock URL
  \url{https://genai.owasp.org/2025/12/09/owasp-genai-security-project-releases-top-10-risks-and-mitigations-for-agentic-ai-security/}.
\newblock Announcement dated 9 December 2025; release dated 10 December 2025.

\bibitem[{OWASP GenAI Security
  Project}(2025{\natexlab{b}})]{owaspAgenticAiThreatsMitigations}
{OWASP GenAI Security Project}.
\newblock Agentic {AI}: Threats and mitigations, 2025{\natexlab{b}}.
\newblock URL
  \url{https://genai.owasp.org/resource/agentic-ai-threats-and-mitigations/}.
\newblock Agentic Security Initiative, 17 February 2025.

\bibitem[Parnas(1972)]{parnas1972criteria}
David~L. Parnas.
\newblock On the criteria to be used in decomposing systems into modules.
\newblock \emph{Communications of the ACM}, 15\penalty0 (12):\penalty0
  1053--1058, 1972.
\newblock \doi{10.1145/361598.361623}.

\bibitem[{Python Software Foundation}(n.d.)]{pythonTyping2026}
{Python Software Foundation}.
\newblock typing: Support for type hints.
\newblock Python documentation, n.d.
\newblock URL \url{https://docs.python.org/3/library/typing.html}.

\bibitem[Rao and Georgeff(1995)]{raoGeorgeff1995bdiAgents}
Anand~S. Rao and Michael~P. Georgeff.
\newblock {BDI} agents: From theory to practice.
\newblock In \emph{Proceedings of the First International Conference on
  Multi-Agent Systems}, pages 312--319, 1995.
\newblock URL \url{https://cdn.aaai.org/ICMAS/1995/ICMAS95-042.pdf}.
\newblock ICMAS-95, San Francisco.

\bibitem[{Restate}(2026)]{restateDurableExecution2026}
{Restate}.
\newblock Key concepts: Durable execution, 2026.
\newblock URL
  \url{https://docs.restate.dev/foundations/key-concepts#durable-execution}.

\bibitem[Sabelfeld and Myers(2003)]{sabelfeld2003informationFlow}
Andrei Sabelfeld and Andrew~C. Myers.
\newblock Language-based information-flow security.
\newblock \emph{IEEE Journal on Selected Areas in Communications}, 21\penalty0
  (1):\penalty0 5--19, 2003.
\newblock \doi{10.1109/JSAC.2002.806121}.
\newblock URL \url{https://www.cs.cornell.edu/andru/papers/jsac/sm-jsac03.pdf}.

\bibitem[Saltzer and Schroeder(1975)]{saltzer1975protection}
Jerome~H. Saltzer and Michael~D. Schroeder.
\newblock The protection of information in computer systems.
\newblock \emph{Proceedings of the IEEE}, 63\penalty0 (9):\penalty0 1278--1308,
  1975.
\newblock \doi{10.1109/PROC.1975.9939}.
\newblock URL
  \url{https://web.mit.edu/Saltzer/www/publications/protection/index.html}.

\bibitem[Schick et~al.(2023)Schick, Dwivedi-Yu, Dess{\`i}, Raileanu, Lomeli,
  Hambro, Zettlemoyer, Cancedda, and Scialom]{schick2023toolformer}
Timo Schick, Jane Dwivedi-Yu, Roberto Dess{\`i}, Roberta Raileanu, Maria
  Lomeli, Eric Hambro, Luke Zettlemoyer, Nicola Cancedda, and Thomas Scialom.
\newblock Toolformer: Language models can teach themselves to use tools.
\newblock In \emph{Advances in Neural Information Processing Systems 36}, 2023.
\newblock \doi{10.52202/075280-2997}.
\newblock URL \url{https://openreview.net/pdf?id=Yacmpz84TH}.
\newblock NeurIPS 2023.

\bibitem[Schneider(2000)]{schneider2000enforceable}
Fred~B. Schneider.
\newblock Enforceable security policies.
\newblock \emph{ACM Transactions on Information and System Security},
  3\penalty0 (1):\penalty0 30--50, 2000.
\newblock \doi{10.1145/353323.353382}.
\newblock URL \url{https://www.cs.cornell.edu/fbs/publications/EnfSecPols.pdf}.

\bibitem[Shi et~al.(2026)Shi, He, Wang, Li, Wu, Guo, and Song]{shi2026progent}
Tianneng Shi, Jingxuan He, Zhun Wang, Hongwei Li, Linyu Wu, Wenbo Guo, and Dawn
  Song.
\newblock {Progent}: Securing {AI} agents with privilege control.
\newblock \emph{arXiv}, 2026.
\newblock \doi{10.48550/arXiv.2504.11703}.
\newblock URL \url{https://arxiv.org/abs/2504.11703v3}.
\newblock arXiv v3, 14 May 2026.

\bibitem[Shoham(1993)]{shoham1993agentOrientedProgramming}
Yoav Shoham.
\newblock Agent-oriented programming.
\newblock \emph{Artificial Intelligence}, 60\penalty0 (1):\penalty0 51--92,
  1993.
\newblock \doi{10.1016/0004-3702(93)90034-9}.

\bibitem[{Sigstore}(2026)]{sigstore2026docs}
{Sigstore}.
\newblock Sigstore documentation, 2026.
\newblock URL \url{https://docs.sigstore.dev/}.

\bibitem[{SLSA}(2026)]{slsa2026spec}
{SLSA}.
\newblock Supply-chain levels for software artifacts specification, 2026.
\newblock URL \url{https://slsa.dev/spec/v1.2/}.
\newblock Version 1.2.

\bibitem[Strom and Yemini(1986)]{strom1986typestate}
Robert~E. Strom and Shaula Yemini.
\newblock Typestate: A programming language concept for enhancing software
  reliability.
\newblock \emph{IEEE Transactions on Software Engineering}, SE-12\penalty0
  (1):\penalty0 157--171, 1986.
\newblock \doi{10.1109/TSE.1986.6312929}.
\newblock URL
  \url{https://research.ibm.com/publications/typestate-a-programming-language-concept-for-enhancing-software-reliability}.

\bibitem[{Temporal Technologies}(2026)]{temporalDurableExecution2026}
{Temporal Technologies}.
\newblock What is {Temporal}?, 2026.
\newblock URL \url{https://docs.temporal.io/temporal}.

\bibitem[{The Rust Project}(n.d.)]{rustOwnership2026}
{The Rust Project}.
\newblock What is ownership?
\newblock The Rust Programming Language, n.d.
\newblock URL
  \url{https://doc.rust-lang.org/book/ch04-01-what-is-ownership.html}.

\bibitem[{Treasury Board of Canada
  Secretariat}(2025{\natexlab{a}})]{canada2025admDirective}
{Treasury Board of Canada Secretariat}.
\newblock Directive on automated decision-making, 2025{\natexlab{a}}.
\newblock URL \url{https://www.tbs-sct.canada.ca/pol/doc-eng.aspx?id=32592}.
\newblock Revised 24 June 2025.

\bibitem[{Treasury Board of Canada
  Secretariat}(2025{\natexlab{b}})]{canada2025admScope}
{Treasury Board of Canada Secretariat}.
\newblock Guide on the scope of the directive on automated decision-making,
  2025{\natexlab{b}}.
\newblock URL
  \url{https://www.canada.ca/en/government/system/digital-government/digital-government-innovations/responsible-use-ai/automated-decision-making/guide-scope-directive-automated-decision-making.html}.

\bibitem[Wang et~al.(2024)Wang, Ma, Feng, Zhang, Yang, Zhang, Chen, Tang, Chen,
  Lin, Zhao, Wei, and Wen]{wang2023llmAgentsSurvey}
Lei Wang, Chen Ma, Xueyang Feng, Zeyu Zhang, Hao Yang, Jingsen Zhang, Zhiyuan
  Chen, Jiakai Tang, Xu~Chen, Yankai Lin, Wayne~Xin Zhao, Zhewei Wei, and
  Ji-Rong Wen.
\newblock A survey on large language model based autonomous agents.
\newblock \emph{Frontiers of Computer Science}, 18\penalty0 (6):\penalty0
  186345, 2024.
\newblock \doi{10.1007/s11704-024-40231-1}.
\newblock URL
  \url{https://journal.hep.com.cn/fcs/EN/10.1007/s11704-024-40231-1}.

\bibitem[{World Health Organization}(2024)]{who2025lmmHealthGuidance}
{World Health Organization}.
\newblock Ethics and governance of artificial intelligence for health: Guidance
  on large multi-modal models.
\newblock Technical report, World Health Organization, 2024.
\newblock URL \url{https://www.who.int/publications/i/item/9789240084759}.

\bibitem[Wu et~al.(2024)Wu, Bansal, Zhang, Wu, Li, Zhu, Jiang, Zhang, Zhang,
  Liu, Awadallah, White, Burger, and Wang]{wu2023autogen}
Qingyun Wu, Gagan Bansal, Jieyu Zhang, Yiran Wu, Beibin Li, Erkang Zhu,
  Li~Jiang, Xiaoyun Zhang, Shaokun Zhang, Jiale Liu, Ahmed~Hassan Awadallah,
  Ryen~W. White, Doug Burger, and Chi Wang.
\newblock {AutoGen}: Enabling next-gen {LLM} applications via multi-agent
  conversations.
\newblock In \emph{First Conference on Language Modeling}, 2024.
\newblock URL \url{https://openreview.net/pdf?id=BAakY1hNKS}.

\bibitem[Yao et~al.(2023)Yao, Zhao, Yu, Du, Shafran, Narasimhan, and
  Cao]{yao2022react}
Shunyu Yao, Jeffrey Zhao, Dian Yu, Nan Du, Izhak Shafran, Karthik Narasimhan,
  and Yuan Cao.
\newblock {ReAct}: Synergizing reasoning and acting in language models.
\newblock In \emph{International Conference on Learning Representations}, 2023.
\newblock URL \url{https://arxiv.org/abs/2210.03629v3}.
\newblock ICLR 2023.

\end{thebibliography}
}

\begin{document}

  \maketitle
  \begin{abstract}
Language-model agents combine reasoning, tools, memory, delegation, and external
effects. Authority is often distributed across prompts, host code, configuration,
and logs. Probabilistic Agentic Programming (PAP) makes responsibility boundaries
explicit programming abstractions connecting mandate, authority, effects, and
evidence.

We compare PAP with programming paradigms, agent-oriented and BDI systems,
workflow and policy languages, and LLM tool-use frameworks along seven dimensions.
A mediation kernel formalizes checks before tool execution, operation-bound
approval, resource limits, and evidence. Inductive proofs establish conditional
safety for arbitrary model responses, assuming correct guards and complete
mediation. A backward simulation matches graph executions to source executions
in a defined sequential Touba subset. Its human checkpoints release program
continuations; operation-bound approval remains a separate kernel contract.
Finite-state exploration checks the model. Scripted enforcement tests exercise
selected refusal paths and permitted controls.

An exploratory study combines seven business cases, including two three-agent
workflows, seven models from three providers, five methods, and five repetitions
per case and method. Shared English instructions and exact checks of stored
outcomes yield 1,225 assessed observations. Native Touba completes all 35
observations on six models and 30 on the remaining model; structured Touba ranges
from 21 to 35. Native matches the highest observed count for each model, including
ties with the frameworks. All paired case intervals include zero. These observed
counts describe the evaluated tasks; the runtime authority boundary remains
fixed. The results connect responsibility contracts, execution rules, and evidence.

\end{abstract}

  \PaperPrintKeywords

\section{Introduction}

Programming paradigms shape what programmers see as a first-class design
object. Structured programming made uncontrolled jumps a design problem;
modular programming made information hiding and change localization central;
functional programming emphasized expression and composition; actors made
active entities and messages explicit~\cite{dijkstra1968goto,parnas1972criteria,backus1978vonNeumann,hewitt1973actors}.

Language-model agents create a different pressure. They can select tools,
consult memory, delegate, request human judgment, and affect external systems
through probabilistic decisions. The facts that govern an action (responsibility,
authority, capability, memory scope, validation, binding, and evidence) can be
scattered across the stack. Probabilistic Agentic Programming (PAP) studies that scattering as a programming
problem: the agent becomes a responsibility-bearing unit whose permitted
effects and obligations are explicit.

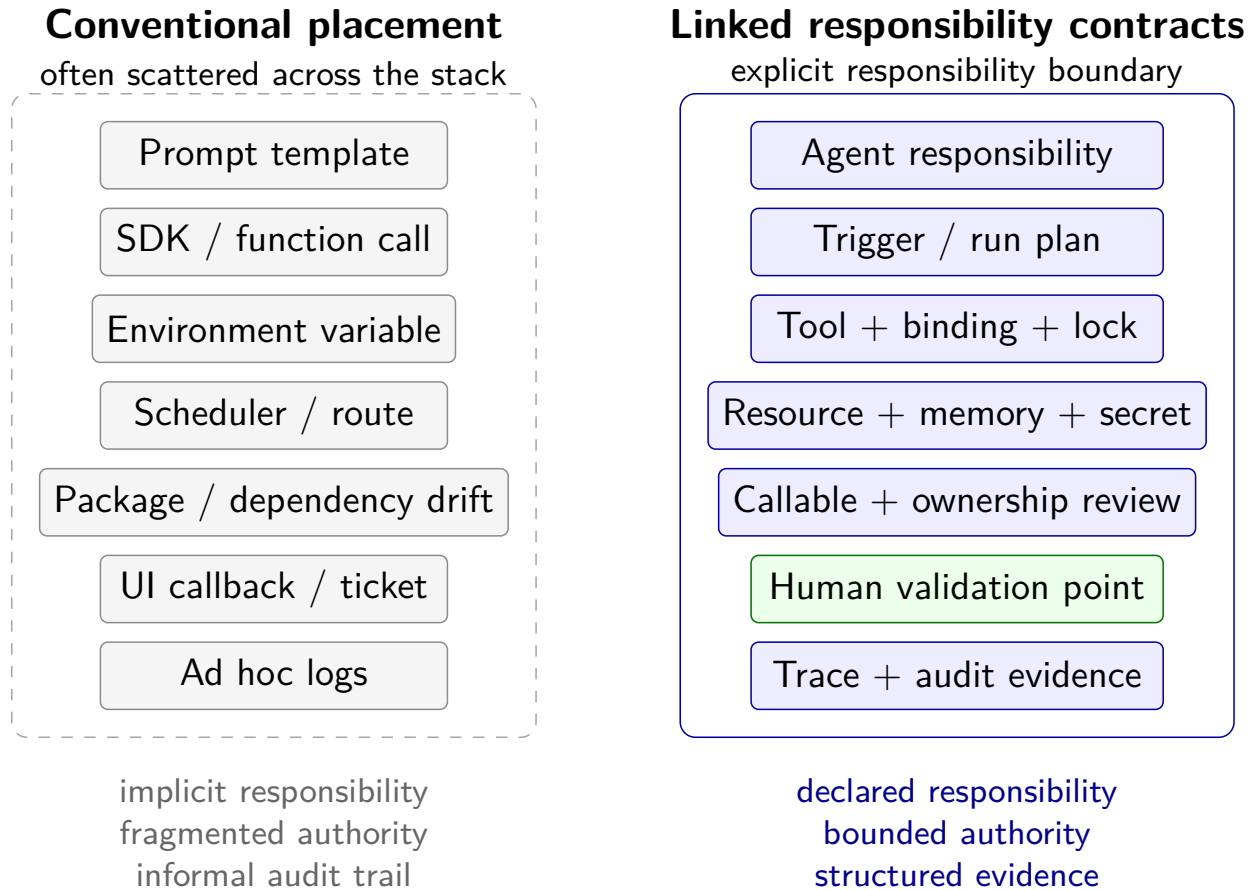
\begin{figure}[htbp]
\centering
\resizebox{\textwidth}{!}{%
\begin{tikzpicture}[
  font=\sffamily\small,
  conventional/.style={draw=black!45, fill=black!4, rounded corners=2pt, align=center, minimum width=3.2cm, minimum height=0.62cm},
  pap/.style={draw=blue!65!black, fill=blue!7, rounded corners=2pt, align=center, minimum width=3.8cm, minimum height=0.62cm},
  human/.style={draw=green!45!black, fill=green!8, rounded corners=2pt, align=center, minimum width=3.8cm, minimum height=0.62cm},
  label/.style={font=\sffamily\bfseries, align=center},
  smalllabel/.style={font=\sffamily\footnotesize, align=center}
]
  \node[label] at (0,0) {Conventional placement};
  \node[smalllabel] at (0,-0.42) {often scattered across the stack};

  \node[conventional] (prompt) at (0,-1.18) {Prompt template};
  \node[conventional] (function) at (0,-1.98) {SDK / function call};
  \node[conventional] (env) at (0,-2.78) {Environment variable};
  \node[conventional] (scheduler) at (0,-3.58) {Scheduler / route};
  \node[conventional] (dependency) at (0,-4.38) {Package / dependency drift};
  \node[conventional] (ui) at (0,-5.18) {UI callback / ticket};
  \node[conventional] (logs) at (0,-5.98) {Ad hoc logs};
  \node[smalllabel, text=black!60] at (0,-7.42) {implicit responsibility\\fragmented authority\\informal audit trail};

  \node[label] at (6.3,0) {Linked responsibility contracts};
  \node[smalllabel] at (6.3,-0.42) {explicit responsibility boundary};

  \node[pap] (agent) at (6.3,-1.18) {Agent responsibility};
  \node[pap] (trigger) at (6.3,-1.98) {Trigger / run plan};
  \node[pap] (tool) at (6.3,-2.78) {Tool + binding + lock};
  \node[pap] (secret) at (6.3,-3.58) {Resource + memory + secret};
  \node[pap] (authority) at (6.3,-4.38) {Callable + ownership review};
  \node[human] (validation) at (6.3,-5.18) {Human validation point};
  \node[pap] (trace) at (6.3,-5.98) {Trace + audit evidence};
  \node[smalllabel, text=blue!55!black] at (6.3,-7.42) {declared responsibility\\bounded authority\\structured evidence};

  \node[draw=blue!65!black, rounded corners=4pt, fit=(agent)(trigger)(tool)(secret)(authority)(validation)(trace), inner sep=0.25cm] {};
  \node[draw=black!35, dashed, rounded corners=4pt, fit=(prompt)(function)(env)(scheduler)(dependency)(ui)(logs), inner sep=0.25cm] {};
\end{tikzpicture}%
}
\caption{Organization of governance mechanisms. PAP links governance-relevant
facts through explicit responsibility contracts.}
\label{fig:pap-conventional-vs-boundary}
\end{figure}

Figure~\ref{fig:pap-conventional-vs-boundary} contrasts this organization with
a conventional distribution of governance mechanisms. PAP connects existing
mechanisms through a common responsibility boundary; it builds on computation,
orchestration, authorization, and established agent traditions.

The founding account introduces PAP as a programming paradigm~\cite{lo2026foundationalPAP}.
This paper develops its comparative account of responsibility boundaries,
a formal preservation result, and an executable evaluation. The comparison
identifies how programming traditions and tool-use stacks represent boundary
obligations. A mediation kernel specifies the checks before execution,
operation-bound human decisions, resource limits, and evidence. A backward
simulation relates graph execution to source contracts in a defined Touba subset.
Compiled tests check selected implementation mechanisms, and business cases
examine completion across model-facing interfaces.

The evaluation links model proposals, runtime events and independently
observed effects. It combines scripted enforcement tests
(Section~\ref{sec:contained-failures}) with case-paired business outcomes
(Section~\ref{sec:benchmark-evidence}). Comparing two Touba decision interfaces
examines completion while runtime authority remains fixed.

\section{Research Question}

This study combines source-backed comparison with executable benchmark evidence. It clarifies the distinct programming concern named by PAP and shows how that concern relates to established paradigms and current agentic tool-use stacks.

\subsection{Primary Question}

The primary question is:

\begin{quote}
\small
How does PAP differ from classical programming paradigms,
historical agent-oriented programming, BDI agents, actor systems, and recent LLM tool-use systems?
\end{quote}

\subsection{Subquestions}

The comparison asks four subquestions: what central abstraction each tradition
makes visible, where responsibility lives, how effects, authority, memory and
validation are represented, and what execution evidence can be audited.
Together, they distinguish a tradition's computational strengths from its
treatment of agentic responsibility.

The paper develops an emerging programming paradigm. Its design claim is that a
responsibility boundary connects facts otherwise represented separately:
mandate, authority, permitted effects and attributable evidence. We evaluate
the obligations of that connection, their formal preservation in a defined
realization, and their representation in executable systems.

\subsection{Boundary of the Claim}

PAP addresses a specific and increasingly common setting: systems where probabilistic agents decide, use tools, consult memory, cross authority boundaries, interact with humans, and produce effects that must be governed and audited. Structured, functional, modular, object-oriented, actor-based, and agent-oriented programming remain the computational foundations around which this responsibility boundary is added.

The comparison asks whether PAP names a distinct programming object: governed responsibility under probabilistic agency.

Here `probabilistic' names the execution condition under study: systems whose behavior may involve stochastic model components, uncertain retrieval, adaptive planning, partial observations, and human judgment. The comparative claim is about governing responsibility around such components and positioning that concern alongside probabilistic programming, POMDPs, reinforcement learning, and Bayesian decision theory.

\section{Background}

\subsection{Classical Programming Paradigms}

Classical paradigms stabilize different design questions. Structured programming emphasizes disciplined control flow and the avoidance of arbitrary jumps~\cite{dijkstra1968goto}. Functional programming emphasizes expression, composition, and a critique of the assignment-centered von Neumann style~\cite{backus1978vonNeumann}. Modular programming emphasizes information hiding and the decomposition of systems around decisions that may change~\cite{parnas1972criteria}. Object-oriented programming organizes programs around objects that combine identity, state, and behavior; its history is broad and cannot be reduced to a single feature~\cite{black2013oopHistory}.

These traditions remain necessary. PAP depends on them. An implementation following PAP still needs functions, types, modules, protocols, tests, and runtime engineering. Probabilistic agentic systems add a design concern that deserves first-class treatment: who is responsible for which autonomous action, under which authority and validation boundary.

\subsection{Actors and Concurrent Agents}

Actor systems model independent computational entities that communicate by message passing. Hewitt, Bishop, and Steiger introduced actors as a modular formalism for artificial intelligence~\cite{hewitt1973actors}. This makes actors one of the closest historical comparisons for the paradigm: from the outside, an agent following PAP can look like an active unit that receives an invocation, performs work, and emits results or messages.

The actor model makes concurrency, communication, and local state management
explicit. PAP adds the responsibility boundary of an active unit that chooses
actions under uncertainty, linking its behavior to enforceable contracts.

\subsection{Agent-Oriented Programming and BDI Agents}

Shoham's Agent-Oriented Programming proposed agents whose behavior is described in mentalistic terms such as beliefs, commitments, capabilities, and decisions~\cite{shoham1993agentOrientedProgramming}. BDI architectures developed a practical agent tradition around beliefs, desires, and intentions~\cite{raoGeorgeff1995bdiAgents}.

2APL makes this tradition concrete through a language with declarative beliefs
and goals, executable plans, and rules for plan generation and
repair~\cite{dastani2008twoAPL}. It provides a useful language-level comparison
for how agent concepts become explicit syntax and operational semantics.

PAP inherits autonomy from these traditions and centers the contracts that
govern its exercise, including when the decision mechanism is a language model.

\subsection{LLM Tool-Use Systems}

Recent systems combine language models with external tools, reasoning steps, retrieval, and multi-agent coordination. MRKL systems propose modular architectures that combine language models, external knowledge sources, and symbolic reasoning~\cite{karpas2022mrklSystems}. ReAct interleaves reasoning and acting in language models~\cite{yao2022react}. Toolformer studies how models can learn to use tools~\cite{schick2023toolformer}. AutoGen explores multi-agent conversation as an application pattern~\cite{wu2023autogen}. Current SDKs and protocols operationalize the same pressure in production-oriented stacks: the OpenAI Agents SDK exposes agents, tools, handoffs, guardrails, sessions, human-in-the-loop, MCP integration, and tracing; Google ADK exposes multi-agent workflows, tool integrations, runtime controls, observability, evaluation, safety, sessions, memory, skills, and context management; and MCP standardizes a protocol boundary for connecting models to external data and tools~\cite{openaiAgentsSdk2026,googleAdkDocs2026,anthropicModelContextProtocol2025}. LangGraph uses persistence and checkpointers to support continuation, human-in-the-loop workflows, time travel, and fault tolerance~\cite{langgraphPersistence2026}. Surveys document a broad and rapidly changing space of LLM-based autonomous agents~\cite{wang2023llmAgentsSurvey}. Agentic security work, from AgentDojo and CaMeL to 2026 dynamic and automated prompt-injection evaluations, shows that tool-use stacks also need adversarial evaluation and explicit data/control boundaries~\cite{agentdojo2024benchmark,camel2025defeatingPromptInjections,agentdyn2026benchmark,hofer2026automatedPromptInjection}.

These works combine reasoning and action. PAP asks how their responsibility
boundaries can become explicit programming constructs.

\subsection{Workflow and Policy Languages}

Workflow standards provide another important comparison point. WS-BPEL specifies executable and abstract business processes based on Web Services, including service interactions and process coordination~\cite{oasis2007wsbpel}. Modern durable-execution systems such as Temporal and Restate make persistence, replay control, and recovery central runtime concerns for long-running workflows~\cite{temporalDurableExecution2026,restateDurableExecution2026}. Their center of gravity is orchestration and durable execution: describing how activities and partner interactions are ordered, recorded, resumed, and coordinated.

Policy languages such as XACML address a different part of the problem. XACML defines a policy language and an access-control decision request/response model for authorization decisions~\cite{oasis2013xacml}. Modern policy systems such as OPA/Rego and Cedar similarly make policy decisions explicit, analyzable, and separable from application code~\cite{opa2026docs,cedar2024authorizationLanguage}. Their center of gravity is policy evaluation: deciding whether a subject may perform an action on a resource under given attributes, obligations, and combining rules.

PAP connects orchestration and policy evaluation to the agent responsible for
an action. The responsibility boundary links these mechanisms within the
program.

\subsection{Language-Based Security and Enforcement}
\label{sec:security-positioning}

PAP's enforcement foundations predate LLM agents. Complete mediation and
fail-safe defaults are classical protection principles~\cite{saltzer1975protection}.
Schneider characterizes the safety policies enforceable by monitoring
executions~\cite{schneider2000enforceable}. The mediation kernel applies this
monitoring tradition to responsibility boundaries: admission connects an
operation to its responsible context and declared contracts.

Here, a \emph{capability} means a callable operation or agent entry exposed for
selection; \emph{authority} determines whether its invocation is permitted.
Object-capability systems organize authority through references and controlled
composition~\cite{miller2006robustComposition}. PAP likewise makes authority
explicit, but a Touba uses declaration is a runtime-checked selection relation,
not an unforgeable object capability. Agent delegation requires both declared
selection and an admitted caller; the specialist retains its own declared
capabilities. The active responsibility and invocation chain identify where
the resulting effect belongs.

Effect systems statically describe computational effects~\cite{lucassen1988effects},
whereas typestate constrains operations according to protocol
state~\cite{strom1986typestate}. Touba's pure-function boundary and the
kernel's pending, waiting, and in-flight phases address related concerns
through explicit contracts and runtime admission. PAP connects those checks
to agent responsibilities and execution evidence.

Information-flow security governs propagation after access has been
granted~\cite{sabelfeld2003informationFlow}. Scoped memory access and secret
handles govern admission; proving noninterference additionally requires
tracking information propagation. Proof-carrying code equips code with a
proof that a consumer can check against a safety
policy~\cite{necula1997proofCarryingCode}. A kernel dispatch witness serves a
different purpose: it records the contract checks justifying one execution step.

Recent agent defenses apply related ideas directly to tool use. CaMeL separates
trusted control flow from untrusted data and tracks dependencies to enforce
data-use policies~\cite{camel2025defeatingPromptInjections}. Progent checks tool
names and arguments against symbolic policies; its revised design checks
policy updates with an SMT solver and requires approval for privilege
expansion~\cite{shi2026progent}. These approaches share the goal of enforcing
boundaries outside the model. PAP centers the programming abstraction that
connects such controls to responsibility, delegation and execution evidence.

These distinctions sharpen the proposed contribution: a responsibility
boundary is a programming abstraction whose identity, authority relations,
and obligations must survive lowering, mediation, and evidence production.
Section~\ref{sec:boundary-preservation} states a bounded preservation theorem
for that chain.

\section{Governance Pressure in Sensitive Organizations}

Public and sectoral guidance motivates explicit responsibility in software.
AI systems affecting health, financial decisions, public services, safety,
or fundamental rights require governance, inventories, testing, supervision,
and accountable operation.

The NIST AI RMF places trustworthiness within the full AI life
cycle~\cite{nist2023aiRmf}; its Generative AI profile identifies privacy,
security, confabulation, human-AI configuration, information-integrity and
value-chain risks~\cite{nist2024genAiProfile}. OWASP examines the attack surfaces
created by agents combining reasoning, tools and
interaction~\cite{owasp2025agenticTop10}. WHO addresses governance of large
multi-modal models in health~\cite{who2025lmmHealthGuidance}, while OSFI connects
model-risk governance and approval roles to institutional risk
appetite~\cite{osfi2025e23}.

Government guidance provides further examples. OMB M-24-10 required federal
AI governance, inventories, and identification of safety- and rights-impacting
uses~\cite{omb2024m2410}; it was rescinded and replaced by M-25-21 on
3 April 2025~\cite{omb2025m2521}. Canada's Directive on Automated Decision-Making
sets requirements for impact assessment, transparency, quality assurance,
human involvement and recourse~\cite{canada2025admDirective}. Its scope guide
explains which administrative uses are covered~\cite{canada2025admScope}.
The EU AI Act establishes a risk-based framework with requirements for
high-risk systems~\cite{eu2024aiAct}.

These sources motivate concrete programming questions: which agent was
authorized to act, which tool produced an effect, which resource scope was
available, which human decision was required, and which evidence reconstructs
the execution while protecting sensitive values?

The programming substrate must also connect to operational infrastructure.
OpenTelemetry provides an observability vocabulary, while SLSA, in-toto and
Sigstore address provenance, attestations and signing~\cite{opentelemetryGenAi2026,slsa2026spec,inToto2026Framework,sigstore2026docs}.
Touba makes selected governance facts inspectable through source, AIR,
bindings and traces. Organizational assurance additionally depends on the
surrounding infrastructure, policies and sector-specific evaluation.

\section{Threat Model and Trust Assumptions}
\label{sec:threat-model}

Model proposals and tool observations may be wrong or adversarial. The
kernel trusts contract evaluators and complete mediation, while the execution
study additionally trusts its adapters, local services and measurement path.
The actual compiler/runtime is a subject of testing, not assumed verified.
Table~\ref{tab:threat-model} separates prevention, observation and threats
requiring additional mechanisms.

\begin{table}[htbp]
\centering
\scriptsize
\setlength{\tabcolsep}{3pt}
\begin{tabular}{@{}>{\raggedright\arraybackslash}p{0.16\linewidth}>{\raggedright\arraybackslash}p{0.25\linewidth}>{\raggedright\arraybackslash}p{0.24\linewidth}>{\raggedright\arraybackslash}p{0.25\linewidth}@{}}
\toprule
Threat & Required condition & Boundary response & Evidence or remaining obligation \\
\midrule
Unlisted tool or invalid caller & Trusted identity and usage/caller guards & Refused before handler dispatch & Kernel proof; compiled delegation and direct-call checks. \\
Missing resource grant or exhausted quota & Correct resource evaluators and counters & Refused at admission & Compiled resource tests and independently observed writes. \\
Missing required human decision & Approval contract defined in Section~\ref{sec:boundary-preservation} & Wait or refuse according to that contract & Kernel proof and compiled checkpoint tests. \\
Injected or false observation & Provider cannot change authority; all effects mediated & An inadmissible proposal is blocked & Business outcomes checked separately by the task oracle. \\
Malicious authorized handler & Handler already admitted & Handler internals are outside the kernel & Requires effect isolation or service-side enforcement. \\
Policy changes between check and effect & Atomic admission inputs in the model & No distributed race guarantee & Requires a service-side authorization/commit protocol. \\
Crash, missing or altered trace & Crash-free execution and lossless abstract audit & No physical completeness or tamper-resistance guarantee & Requires durable correlation and evidence integrity. \\
Concurrent or distributed delegation & Sequential verified contexts & Outside the preservation fragment & Requires a model of shared authority and concurrent effects. \\
\bottomrule
\end{tabular}
\caption{Threats, conditional boundary responses and supporting evidence.}
\label{tab:threat-model}
\end{table}

Arbitrary provider responses are covered by the kernel argument; compromise
of the trusted execution host or its credentials is not. Independent reads
of instrumented service state strengthen the selected execution checks,
but do not make the measurement infrastructure adversary-resistant.

\section{Comparison Dimensions}

A comparison between PAP and earlier paradigms needs dimensions that are specific enough to be falsifiable. This paper uses seven dimensions.

The first dimension is the central abstraction: the principal unit that programmers reason about. The next two dimensions ask where responsibility lives and how calls to the external world are represented. Authority then asks whether invocation rights are explicit, implicit, host-language based, or policy-mediated. Memory asks whether context is ordinary state, agent belief, shared context, scoped resource, or external store. Human validation asks whether judgment is a built-in design object, an external workflow, or an application convention. Execution evidence asks whether the system can produce structured evidence linked to source-level responsibilities, or only logs and traces at the host level.

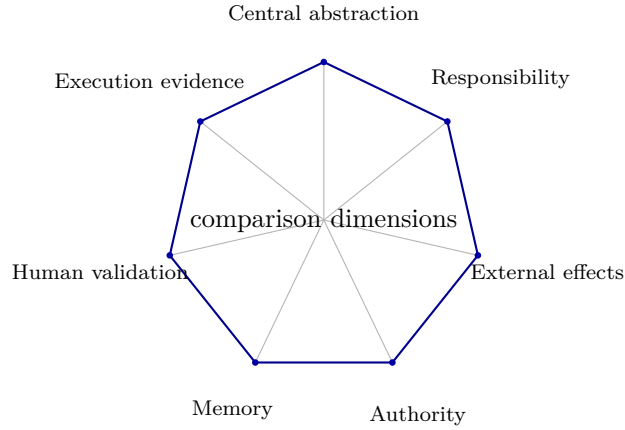
\begin{figure}[htbp]
\centering
\begin{tikzpicture}[scale=0.82, every node/.style={font=\scriptsize}]
  \draw[blue!55!black, thick]
    (90:2.55) -- (38.6:2.55) -- (-12.9:2.55) -- (-64.3:2.55) --
    (-115.7:2.55) -- (-167.1:2.55) -- (141.4:2.55) -- cycle;
  \foreach \angle in {90,38.6,-12.9,-64.3,-115.7,-167.1,141.4} {
    \draw[gray!60] (0,0) -- (\angle:2.55);
    \fill[blue!65!black] (\angle:2.55) circle (1.5pt);
  }
  \node[align=center] at (90:3.35) {Central\ abstraction};
  \node[align=center] at (38.6:3.65) {Responsibility};
  \node[align=center] at (-12.9:3.7) {External\ effects};
  \node[align=center] at (-64.3:3.5) {Authority};
  \node[align=center] at (-115.7:3.4) {Memory};
  \node[align=center] at (-167.1:3.7) {Human\ validation};
  \node[align=center] at (141.4:3.6) {Execution\ evidence};
  \node[align=center, font=\small] at (0,0) {comparison\ dimensions};
\end{tikzpicture}
\caption{Seven conceptual comparison dimensions. The wheel has no magnitude,
area, or score semantics.}
\label{fig:pap-comparison-radar}
\end{figure}

The wheel organizes the seven qualitative dimensions. It explains why the
comparison covers more than tool support or orchestration: a system may have
strong tooling while placing responsibility, validation, or evidence in a
different layer.

\begin{figure}[htbp]
\centering
\resizebox{0.96\linewidth}{!}{%
\begin{tikzpicture}[
  source/.style={
    draw=black!38,
    fill=black!2,
    rounded corners=3pt,
    align=center,
    text width=3.7cm,
    minimum height=1.05cm,
    inner sep=5pt,
    font=\sffamily\footnotesize
  },
  pap/.style={
    draw=blue!68!black,
    fill=blue!7,
    rounded corners=4pt,
    align=center,
    text width=5.3cm,
    minimum height=1.25cm,
    inner sep=7pt,
    font=\sffamily\small
  },
  arrow/.style={
    -{Latex[length=2.2mm,width=1.6mm]},
    draw=blue!58!black,
    semithick,
    shorten >=2pt,
    shorten <=2pt
  }
]
  \node[source] (programming) at (-4.55,1.45)
    {Programming paradigms\\control, composition,\\modules, objects};
  \node[source] (actors) at (-4.55,-0.55)
    {Actors and\\distributed systems\\active units, messages,\\durability};
  \node[source] (agents) at (0,2.65)
    {Agent-oriented and BDI\\autonomy, beliefs,\\intentions};
  \node[source] (llm) at (4.55,1.45)
    {LLM tool-use stacks\\reasoning, tools, handoffs};
  \node[source] (governance) at (4.55,-0.55)
    {Workflow and policy systems\\process, authorization,\\oversight};

  \node[pap] (pap) at (0,-2.75)
    {\textbf{Probabilistic Agentic\\Programming}\\responsibility boundary for governed\\probabilistic agentic systems};

  \draw[arrow] (programming.south east) -- (pap.105);
  \draw[arrow] (actors.south) -- (pap.130);
  \draw[arrow] (agents.south) -- (pap.north);
  \draw[arrow] (llm.south west) -- (pap.75);
  \draw[arrow] (governance.south) -- (pap.50);
\end{tikzpicture}%
}
\caption{Several traditions converge on the responsibility-boundary problem.
Each tradition contributes a complementary design concern.}
\label{fig:positioning-map}
\end{figure}
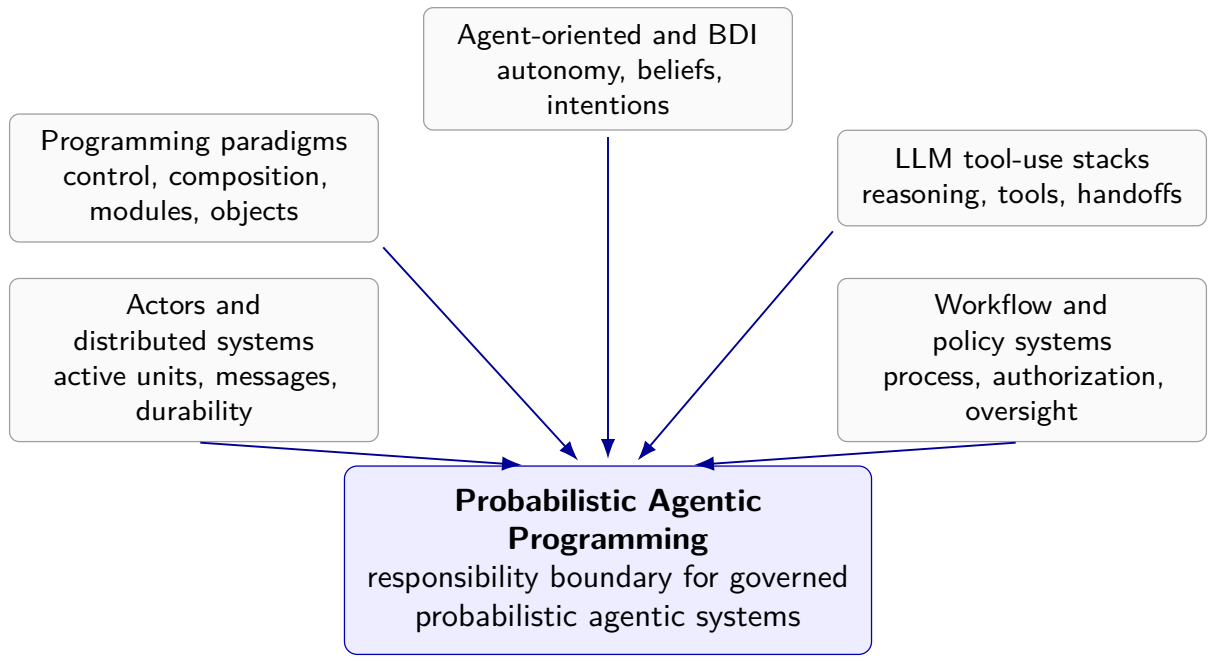

The positioning map shows converging influences rather than a linear historical
succession. Structured computation, actors, AOP/BDI, LLM tools, workflows, and
policy systems remain useful; the paradigm adds a responsibility-oriented lens for
governed probabilistic agency.

These dimensions are chosen for agentic systems. Ordinary criteria such as expressiveness, performance, modularity, type safety, and usability remain important; the selected dimensions isolate the concern that motivates the paradigm: probabilistic autonomy must be programmed with explicit boundaries.

\section{Responsibility as a Programming Abstraction}

PAP organizes probabilistic agentic systems around explicit responsibility
boundaries, connecting permitted action to its authority and evidence.

\subsection{From Computation to Governed Agency}

Classical paradigms often begin with computation: expressions, procedures,
modules, objects, messages, or services. PAP begins with governed agency:
the program exposes who may act and the contracts governing that action.

Ordinary deterministic computation remains central for transformations,
formatting, checks, and internal logic. PAP distinguishes this computation
from governed effects. Touba realizes the distinction through pure functions
and explicit tool contracts.

\subsection{From Agent as Metaphor to Agent as Responsibility Boundary}

Historical agent-oriented programming often models agents through mentalistic notions such as beliefs and commitments~\cite{shoham1993agentOrientedProgramming}. BDI architectures make beliefs, desires, and intentions central~\cite{raoGeorgeff1995bdiAgents}. PAP builds on these traditions and changes the first question asked by the programming model.

The first question is: \emph{what is this agent allowed to do, on whose authority, with which skills, through which tools, using which memory, under which human validation obligations, and with what trace evidence?}

The distinctive object is the connection among these facts. A realization following PAP
identifies the responsible context, separates proposals from authority,
mediates effects and delegation against declared contracts, respects resource
and human obligations, and retains attributable evidence. Composition must
preserve these connections: a tool policy must be evaluated under the correct
agent identity, and its decision must remain attributable to that context.
Capabilities, policies and logs can realize this abstraction in different
languages. PAP specifies these architectural obligations independently of
Touba's syntax.

Responsibility has three complementary aspects, distinguished in
Table~\ref{tab:responsibility-aspects}: mandate, executable contracts, and
action attribution. Here responsibility has this programming meaning;
legal accountability is an organizational question.

\begin{table}[htbp]
\centering
\small
\begin{tabular}{@{}>{\raggedright\arraybackslash}p{.22\linewidth}
                  >{\raggedright\arraybackslash}p{\dimexpr.78\linewidth-2\tabcolsep\relax}@{}}
\toprule
Aspect & Operational meaning \\
\midrule
Declared mandate & States the intended task. Automatically checking its fulfilment requires explicit business criteria beyond the natural-language description. \\
Executable contracts & Determine permitted capabilities, callers, resources and required review. These conditions govern whether an operation may start. \\
Action attribution & Connects a proposal, authorization decision and effect to the responsible context and source contracts. \\
\bottomrule
\end{tabular}
\caption{Responsibility as intent, enforceable constraints and attributable action.}
\label{tab:responsibility-aspects}
\end{table}

In Touba, \texttt{responsibility} supplies the mandate, while
\texttt{uses}, \texttt{callable\_by}, resource declarations and host contracts
provide executable constraints. Trace references support attribution.
Passing these checks establishes compliance with the executable constraints.
Fulfilment of the mandate is a separate business question.

\subsection{Autonomy, Human Decisions and Evidence}

Autonomy permits choice within the declared boundary.
Mandatory human approval is imposed by the program or governing authority;
agent-initiated review is an escalation chosen within delegated autonomy.
Their different sources of authority must remain explicit, irrespective of
the mechanism used to implement pause and resume.

PAP requires evidence that attributes permitted and refused actions, resource
access and human decisions to their declared responsibilities, with controlled
disclosure of sensitive information. Structured
logging is one possible substrate; the obligation concerns its connection to
authority and effects. Touba uses source/AIR, caller and binding references
with resource and pause/resume events to realize that connection.

\section{A Formal Responsibility Boundary}
\label{sec:formal-boundary}

A small mediation kernel gives responsibility-boundary obligations precise
execution rules. It governs \emph{dispatch}, the step that starts a tool
operation. A separate program engine tracks call frames and the code to
execute next. The kernel's safety argument holds for arbitrary provider
responses under explicit trust assumptions.

The \emph{paradigm-level requirements} stated in the preceding section guide
the model. \emph{Kernel guarantees} specify justified dispatch, human-boundary
safety, bounded accounting, and ideal audit correspondence. Under the trust
assumptions in Section~\ref{sec:threat-model}, the model uses fixed contracts,
sequential crash-free execution, and lossless abstract auditing. The eight guards and atomic
dispatch are modeling choices. \emph{Touba-specific mechanisms} include
\texttt{responsibility}, \texttt{uses}, \texttt{callable\_by}, AIR lowering,
scope resolution, and checkpoint resumption. The preservation theorem
connects this realization to the kernel.

\subsection{Contracts and Guarded Dispatch}

Let \(C=(\Gamma,\Delta)\) be fixed contracts: \(\Gamma\) contains source
responsibilities, interfaces, use and caller relations, and declared control
flow; \(\Delta\) supplies host bindings, grants, and resource contracts.
A prepared operation is \(r=(u,K,\theta,x,b,p)\), with a fresh occurrence \(u\),
verified execution context \(K\), tool \(\theta\), arguments \(x\), binding \(b\),
and parent reference \(p\). The program engine establishes the responsible
agent, caller and binding.

Let \(S\) denote the kernel state, \(q\) its phase, and \(S'\) the state
after a transition. Table~\ref{tab:kernel-notation} summarizes the notation;
Appendix~\ref{app:boundary-proofs} defines the full state and its rules.

\begin{table}[htbp]
\centering
\small
\renewcommand{\arraystretch}{1.12}
\begin{tabular}{@{}>{\raggedright\arraybackslash}p{.27\linewidth}
                  >{\raggedright\arraybackslash}p{\dimexpr.73\linewidth-2\tabcolsep\relax}@{}}
\toprule
\textbf{Notation} & \textbf{Meaning in the kernel} \\
\midrule
\(C=(\Gamma,\Delta)\) & Fixed source contracts \(\Gamma\) and host contracts \(\Delta\). \\
\(S,q,S'\) & Current state, its control phase, and successor state. \\
\(r,u\) & Prepared operation and its fresh occurrence identifier. \\
\(K,\theta,x,b,p\) & Verified context, target tool, argument values, host binding, and parent reference carried by \(r\). \\
\(J,g_j(C,r)\) & Guard names and the Boolean check indexed by \(j\). \\
\(\mathsf{NeedHuman}(C,r),\chi\) & Whether human approval is required, and whether a valid favorable decision for this occurrence has been received. Both are Boolean. \\
\(\mathit{active}\) & Boolean flag indicating that execution remains active. \\
\(n_p,n_t;\ L_p,L_t\) & Consumed provider/tool calls and their limits for this kernel instance. Subscripts \(p,t\) label the two counter categories. \\
\(G(C,S)\) & Combined Boolean condition permitting dispatch. \\
\(W(C,S),w\) & Witness constructor and the resulting record justifying this dispatch from its pre-state. \\
\bottomrule
\end{tabular}
\caption{Symbols used in the dispatch conditions and transition rule.}
\label{tab:kernel-notation}
\end{table}
\FloatBarrier

Write \(J=\{\mathsf{ctx},\mathsf{use},\mathsf{caller},\mathsf{type},
\mathsf{bind},\mathsf{grant},\mathsf{scope},\mathsf{integrity}\}\).
The named predicates \(g_j(C,r)\), for \(j\in J\), check verified context,
declared usage, invocation authority, argument types, binding resolution,
host grants, resource scopes, and required integrity, respectively.
Missing or unresolved requirements block execution: this is the
\emph{fail-closed} rule. These checks form the operation's \emph{admission}
conditions.
Trusted host facts used by a guard are implicit admission-step parameters,
not additions to the fixed contracts. All guards and the witness use the
same admission inputs for an atomic dispatch; Appendix~\ref{app:preservation}
makes the host input \(\omega\) explicit for resources.
Using the Boolean flags and counters in Table~\ref{tab:kernel-notation}, define
\begin{equation}
\begin{aligned}
G(C,S)={}&\left[\bigwedge_{j\in J}g_j(C,r)\right]
          \land\left[\neg\mathsf{NeedHuman}(C,r)\lor\chi\right]\\
         &{}\land\mathit{active}\land\left[n_t<L_t\right].
\end{aligned}
\label{eq:boundary-guard}
\end{equation}
The operators \(\land\), \(\lor\), and \(\neg\) denote conjunction,
disjunction, and negation, respectively. The generalized conjunction requires
every indexed guard to hold. Thus dispatch
requires every contract check, approval when required, an active execution,
and remaining tool-call quota. Square brackets group Boolean conditions.
The authorization decision and dispatch are atomic in this model, not a
distributed transaction with the external service.

For a pending operation, the positive dispatch rule is
\begin{equation}
\frac{q=\mathsf{pending}\quad G(C,S)\quad w=W(C,S)}
     {C\vdash S\xrightarrow{\mathsf{dispatch}(u,\theta,b,w)}S'}.
\label{eq:boundary-dispatch}
\end{equation}
The bar separates premises from their conclusion: if all conditions above
hold, the transition below is allowed. The judgment \(C\vdash\) means
``under contracts \(C\),'' and the labeled arrow records the dispatch event
while moving from \(S\) to \(S'\). The successor enters the in-flight phase,
increments the tool counter, and appends the event. A pending dispatch
attempt that fails its admission conditions is refused.
The witness \(W(C,S)\) copies the immutable request, contract references,
guard results, approval state, and pre-dispatch counter from the same
pre-state. Validity requires both this exact correspondence and
Eq.~\ref{eq:boundary-guard}; it is not a provider-supplied validity flag.
A public trace need not publish the arguments.

A skill is not a mandatory intermediary. Direct calls, composed calls, and
delegation can reach this boundary. A delegation from \(a_1\) to \(a_2\)
requires both \(a_2\in\mathrm{Uses}_{\Gamma}(a_1)\) and
\(a_1\in\mathrm{Callers}_{\Gamma}(a_2)\), not inclusion of the specialist's
capabilities in the coordinator's capabilities. Here \(\mathrm{Uses}_{\Gamma}(a)\)
is agent \(a\)'s declared capability set, and \(\mathrm{Callers}_{\Gamma}(a)\)
its admitted callers; \(\in\) denotes set membership. Multiple operations of one composition
have distinct occurrences, and repeated uses of one tool remain possible.
A path without an active agent retains a structural owner instead of
inventing an agent identity.

This realization checks an operation against the active specialist's authority.
It does not intersect that authority with the coordinator's or the full call
chain. Admitting a caller therefore authorizes indirect access to the
specialist's declared services. A compromised coordinator can exploit overly
broad specialist contracts through otherwise valid arguments (a confused-deputy
risk~\cite{hardy1988confusedDeputy}). Caller-specific argument policies, narrower specialist interfaces and
service-side checks address that risk. Resource scopes govern access;
information-flow controls would additionally be needed to restrict what a
specialist may return to its caller.

\subsection{Established Kernel Properties}

Under the assumptions above, Appendix~\ref{app:boundary-proofs} establishes
the following properties by induction over the closed transition relation.

\noindent\textbf{Dispatch justification and contract immutability.}\quad
Every dispatch in a finite execution has a witness valid in its immediately
preceding state, and provider responses cannot alter the contract environment.

\noindent\textbf{Occurrence and human-boundary safety.}\quad
A denied occurrence cannot subsequently be dispatched. A required human
boundary cannot be crossed without a correlated favorable decision.
An occurrence has at most one dispatch in this crash-free kernel; this is
neither exactly-once external execution nor a ban on repeated capability use.

\noindent\textbf{Accounting and terminal decisions.}\quad
Exact counters remain within their limits. Provider admission precedes the
request itself. A terminal request is unique for the kernel instance and
cannot authorize another capability or review, even if the provider ignores
the terminal catalogue.

\noindent\textbf{Evidence and permitted paths.}\quad
The ideal audit is the sequence of transition labels, with a corresponding
public projection. Permitted paths include successful dispatch and repeated
calls under fresh occurrences. Durable trace completeness and termination
with an unresponsive provider or handler require additional mechanisms.

These properties hold pathwise, for every admitted provider response.
The justified-dispatch proof is an invariant of the guarded transition rules:
it establishes what a conforming monitor enforces.

\subsection{Backward Simulation of a Sequential Fragment}
\label{sec:boundary-preservation}

Preservation connects source declarations to the context checked at admission.
We define a Touba fragment \(F_T\), a lowering \(L_T\), and paired source/graph
engines in Appendix~\ref{app:preservation}. Delegation installs the specialist's
context and restores the parent on return. Imports preserve resource identity
without adding grants. Resumption retains counters and continuations without
replaying effects.
The fragment covers typed scalar and structured values, pure expressions and
functions, branches and loops, sequential skills and strategies, resolved
local imports, direct tool selection, and agent delegation. Resource guards
cover memory access pairs and secret handles. Mandatory approval and explicit
review statements suspend and resume the program with its continuation,
loop cursors, and counters intact. Repeated capability use remains permitted.
Recursion, recovery, free-form interaction, provider-initiated review,
terminal provider mode, concurrency, and crashes remain outside this fragment.

Touba's source checkpoint releases a saved program continuation and consumes
the supplied decision. Revisiting the same static site requires another
decision. This program-level release is distinct from the kernel's HUMAN
rule, which correlates consent with a particular prepared tool operation.
In the defined Touba realization, \(\mathsf{NeedHuman}(C,r)=false\):
tool admission uses the other kernel guards, and human statements are handled
by the program engine. The operation-bound HUMAN contract is modeled and
proved at kernel level; Touba's checkpoints and the business study's simulated
approver implement different contracts.
Full business consent adds separate contracts: authenticated approvers, an
explicit scope of consent, and checks that the intended action matches that
scope.

For example, the following strategy-body excerpt places review before each
write. Its surrounding strategy takes a list of notes and declares
\texttt{Write}, a tool from \texttt{Text} to \texttt{Text}.

\begin{ToubaCode}[title={A checkpoint site revisited with different arguments},breakable=false]
for note in notes
    human.approve
        with context as note
        into approval
    Write
        with note as note
        into stored
\end{ToubaCode}

With notes \texttt{INV-042} and \texttt{INV-043}, execution first waits with
no write. Approving and resuming writes only the first note, then pauses
at the second visit. A second supplied decision releases the second write.
Rejection at either visit preserves the already completed prefix and prevents
the pending write. The compiled checks verify each pause, decision, handler
count, and journal state.

\noindent\textbf{Theorem 1 (Responsibility-Preserving Backward Simulation for \(F_T\)).}\quad
Let \(P\in F_T\) be well formed, with fixed host contracts \(\Delta\)
and paired provider, human, and host inputs.
Under the kernel assumptions, every finite execution of the graph semantics
of \(L_T(P)\) has a matching source execution, up to administrative steps and
renaming of identifiers. Corresponding dispatches have the same source
responsibility, caller chain, selected capability, binding and grants,
argument values, resource-scope decisions, and pre-dispatch counters.
Corresponding human checkpoints preserve the same waiting and continuation
state. Every dispatch has a valid kernel witness, and their ideal evidence
projections agree.

The direction is graph to source: every finite behavior of the lowered graph
is matched by the defined source semantics. This preserves safety and visible
evidence; progress and termination require separate arguments. The lowering
\(L_T\) is a mathematical construction, not the production compiler.
This use of related transition systems to establish behavior inclusion belongs
to the simulation-proof tradition~\cite{lynch1995simulations}.

The proof establishes contract preservation by structural induction,
relates source and graph continuations and call frames, and applies the
kernel propositions. Three evidence levels follow: kernel proofs justify
dispatch; the preservation theorem connects dispatch to its source context
within the defined semantics; compiled tests assess selected correspondences
with the implementation. Section~\ref{sec:validity} states the verification
needed to extend this result to the full compiler and runtime.

\subsection{Executable Validation of the Kernel}

The exploration checks \(2^8\times 2\times 3\times 3=4{,}608\) contract
\emph{configurations}: eight Boolean guards, human approval required or not,
and provider/tool quotas independently ranging from zero to two. Each
configuration is explored with at most two operation occurrences and two
abstract request symbols. A \emph{control state} records the execution phase,
request, approval status, and counters; a \emph{transition} is one enabled
step, such as approval, dispatch, or refusal. The totals, 203,712 reachable
states and 314,904 transitions, sum the explorations across configurations,
with states deduplicated per configuration. They quantify coverage of
authority, human-boundary, quota, and audit checks in this finite kernel model.

Histories are omitted from the state key because enabledness
does not consult them; audit extension and dispatch invariants are checked on
each generated transition before merging. All ten deliberately weakened
variants (one per omitted predicate, plus
human-check and quota omissions) are rejected by the invariant checker.
These mutations check sensitivity to the stated invariants. Exploration
checks the finite transition model against those invariants; whether the
specification captures the intended security policy requires separate review.

Separate execution checks use compiled source, the actual runtime, and
isolated local writes. They confirm that one skill selection can perform two
tool calls, that a repeated selection can perform four writes, and that a
quota can block the fourth write without erasing the preceding three.
They also exercise source-to-intermediate-representation facts, refusal of
an unlisted direct tool, provider admission, and typed completion. Additional
compiled checks combine approval and delegation, memory and secret grants,
loop suspension and resumption, early returns, record construction,
short-circuit evaluation, and local module assembly. Negative paths confirm
that approval adds no resource grant and importing a tool adds no authority
to access its memory.
Section~\ref{sec:artifact-validation} places these checks within the wider
artifact validation, while Section~\ref{sec:validity} states the remaining
implementation and assurance boundaries.

\section{Risk-Governance Fit}

OWASP agentic guidance highlights risks arising from planning, tools,
identity, memory and interaction~\cite{owaspAgenticAiThreatsMitigations,owasp2025agenticTop10}.
NIST places these concerns within life-cycle risk management and
measurement~\cite{nist2023aiRmf,nist2024genAiProfile}. PAP translates part of
this pressure into programming obligations: escalation motivates explicit
caller and capability boundaries; data exposure motivates scoped resources
and protected secret handles; audit gaps motivate attributable effects.
Table~\ref{tab:threat-model} identifies which threats the chosen model
addresses, and Section~\ref{sec:contained-failures} demonstrates selected
admission boundaries in execution.

The same design question extends beyond model-selected actions. Triggered
automation needs an identifiable entry contract, limits and replay handling.
Handler integration needs binding identity and integrity checks. Organizational
ownership and review metadata should support accountability without silently
granting execution authority. Touba's broader artifact coverage is summarized
in Section~\ref{sec:artifact-validation}.

\section{Comparative Positioning}

Table~\ref{tab:comparison-positioning} relates each tradition's central
abstraction to the responsibility-boundary question PAP adds.

\begin{table}[htbp]
\centering
\scriptsize
\setlength{\tabcolsep}{3pt}
\begin{tabular}{@{}>{\raggedright\arraybackslash}p{0.20\linewidth}>{\raggedright\arraybackslash}p{0.27\linewidth}>{\raggedright\arraybackslash}p{0.43\linewidth}@{}}
\toprule
Tradition & Central abstraction & Responsibility-boundary question \\
\midrule
Structured, functional, modular, and object-oriented programming & Control flow, expressions, interfaces, objects & Which probabilistic action belongs to which declared responsibility, effect, authority, and evidence boundary? \\
Actors & Active entities and messages & Which authority, capability, validation, and trace obligations govern an active unit? \\
Agent-oriented programming and BDI & Agents, beliefs, desires, intentions, commitments & How is an agent's autonomy bounded by executable responsibility and authority contracts? \\
LLM tool-use stacks and protocols & Model-driven tool selection, handoffs, sessions, orchestration & Where are tool, secret, memory, review, and evidence boundaries represented and enforced? \\
Workflow and durable execution & Ordered, persistent, resumable activities & Who is responsible for probabilistic choices and effects inside the durable process? \\
Policy languages & Authorization decisions over subjects, actions, and resources & How does authorization combine with execution, validation, memory, and evidence? \\
AI governance frameworks & Risk treatment, oversight, inventories, accountability & Which programmable artifacts can provide evidence for organizational judgment? \\
\bottomrule
\end{tabular}
\caption{The responsibility-boundary question across adjacent programming and governance traditions.}
\label{tab:comparison-positioning}
\end{table}

The LLM tool-use row includes MRKL, ReAct, Toolformer, AutoGen, the OpenAI
Agents SDK, MCP, and LangGraph persistence~\cite{karpas2022mrklSystems,yao2022react,schick2023toolformer,wu2023autogen,openaiAgentsSdk2026,anthropicModelContextProtocol2025,langgraphPersistence2026}.
These systems support governance mechanisms. PAP connects those mechanisms
to an explicit responsibility in the program: authority,
effects and evidence belong to the same declared boundary.

How a mechanism is represented and how it is verified are separate questions.
It may be built in, configured, or written in application code; its supporting
evidence may come from documentation, source inspection, or execution tests.
A framework can offer mechanisms beyond those used by the evaluated adapter.

\subsection{Architectural Obligations, Independent of Syntax}

Table~\ref{tab:pap-realizations} separates architectural obligations from their
expression in a language. The Python and Rust columns outline possible
implementation designs. A host-language realization preserving these
obligations would also instantiate the paradigm.

\begin{table}[htbp]
\centering
\scriptsize
\renewcommand{\arraystretch}{1.16}
\setlength{\tabcolsep}{3pt}
\begin{tabular}{@{}>{\raggedright\arraybackslash}p{.23\linewidth}
                  >{\raggedright\arraybackslash}p{.25\linewidth}
                  >{\raggedright\arraybackslash}p{.25\linewidth}
                  >{\raggedright\arraybackslash}p{\dimexpr.27\linewidth-6\tabcolsep\relax}@{}}
\toprule
\textbf{Architectural obligation} & \textbf{Touba realization} & \textbf{Possible Python design} & \textbf{Possible Rust design} \\
\midrule
Authority comes from verified contracts. & Source responsibility, uses and caller relations; runtime context. & Explicit policy records and a dispatcher that establishes the caller. & Encapsulated capability values and a dispatcher that establishes the caller. \\
\addlinespace
Effects require permission checks. & Binding, grant, scope and budget checks before handler dispatch. & Broker checks policy and counters before invoking a callable. & Controlled effect API checks permissions and counters before invoking a handler. \\
\addlinespace
Human decisions satisfy declared obligations. & Source checkpoint, waiting state and checked continuation. & Explicit approval state and checked resumption. & Approval and continuation state types plus runtime decision checks. \\
\addlinespace
Effects and refusals remain attributable. & Source/AIR references, runtime identities and structured trace. & Correlated responsibility, policy and invocation records. & Correlated responsibility, capability and invocation records. \\
\bottomrule
\end{tabular}
\caption{Architectural obligations and implementation choices. The host-language
columns outline designs for future implementation and evaluation.}
\label{tab:pap-realizations}
\end{table}

Python's type annotations are not enforced by its runtime, whereas Rust's
ownership rules constrain memory management at compilation~\cite{pythonTyping2026,rustOwnership2026}.
Neither property alone decides whether an agent may register an invoice.
The operational question is whether the selected architecture
prevents an inadmissible proposal from reaching its handler, including through
delegation and resumption. Section~\ref{sec:contained-failures} exhibits those
points in Touba. Its structural contribution is to preserve responsibility
and authority across source, AIR, admission and evidence as linked artifacts.

\subsection{Representation in the Evaluated Adapters}
\label{sec:adapter-audit}

Table~\ref{tab:adapter-audit} reports a bounded source audit of the adapters
used for the business cases. Installed packages inspected on 9 October 2026
were OpenAI Agents SDK 0.11.6, LangChain 1.5.0 and LangGraph 1.4.4.
The execution campaign records these same installed package versions alongside
its model profiles and source snapshot.
For all three framework adapters, application code selects each stage's tools
and sequences the specialists. It also validates handoffs and shares
provider/tool counters. Native agent execution thus combines with
host-programmed orchestration and limits.

\begin{table}[htbp]
\centering
\scriptsize
\renewcommand{\arraystretch}{1.12}
\setlength{\tabcolsep}{3pt}
\begin{tabular}{@{}>{\raggedright\arraybackslash}p{.17\linewidth}
                  >{\raggedright\arraybackslash}p{.28\linewidth}
                  >{\raggedright\arraybackslash}p{.28\linewidth}
                  >{\raggedright\arraybackslash}p{\dimexpr.27\linewidth-6\tabcolsep\relax}@{}}
\toprule
System & Capability and delegation representation & Human/resource representation & Evidence used here \\
\midrule
Touba & Source interfaces, \texttt{uses}, \texttt{callable\_by}, strategy and AIR; runtime admission. & Source checkpoints and memory/secret contracts; shared service approval and records in the business study. & Compiled contract tests, runtime traces and independent service-state checks. \\
OpenAI Agents SDK & Configured agent tools and output type; adapter sequences stages and validates outputs. & Tool approval is available in the SDK; adapter uses the shared service, not native review. & Source inspection and scripted adapter tests; SDK tracing disabled, model/service telemetry retained. \\
LangChain & \texttt{createAgent}, tool schemas and structured output; adapter sequences stages. & Review middleware is available; adapter configures neither it nor a checkpointer. & Source inspection and scripted adapter tests; captured exchanges and service state. \\
LangGraph & Prebuilt ReAct agent and tool binding; adapter sequences stages and validates outputs. & Interrupt/resume and checkpointing are available; adapter does not configure them. & Source inspection and scripted adapter tests; captured exchanges and service state. \\
\bottomrule
\end{tabular}
\caption{Representation in the inspected adapters. Source inspection documents
the configuration of governance mechanisms; the business oracle evaluates
task completion rather than native review or resource isolation.}
\label{tab:adapter-audit}
\end{table}

Official documentation describes tool-approval interruptions and resumable
state in OpenAI Agents SDK, review middleware in LangChain, and interrupts
with checkpointing in LangGraph~\cite{openaiSdkReview2026,langchainJsReview2026,langgraphJsInterrupts2026}.
OpenAI also provides built-in tracing~\cite{openaiSdkTracing2026}; this adapter
disables it to limit external data transfers. These facilities provide building blocks
for governed applications. The comparison locates the responsibility contracts
within each evaluated configuration. Comparing protection or programming
effort would require matched implementations and dedicated measures.

\section{Touba as an Empirical Reference Path}
\label{sec:touba-reference}

Touba, Thinking Orchestration Using Bounded Agency, realizes PAP through
source constructs for pure computation, tools, skills, agents and strategies.
AIR connects their contracts to host bindings, runtime admission and traces.
The following example tests that connection in the actual toolchain;
broader resource, review and supply-chain coverage is summarized in
Section~\ref{sec:artifact-validation}.

The runtime enforces authority independently of the model-facing format.
The SDK's default interface asks for a structured decision object. Its opt-in
native interface presents the currently authorized capabilities as function
descriptions. The provider proposes a function name and arguments; the
runtime checks that proposal against the current catalogue, source declarations, types,
budgets, delegation, secrets, reviews, effect identity, and replay state before
any dispatch. Initial orientation and final answers also use typed decisions.

\subsection{A Worked Responsibility Boundary}
\label{sec:worked-boundary}

Consider an invoice intake service. An \texttt{InvoiceDesk} coordinates
registration; an \texttt{InvoiceClerk} records the supplied identifier through
\texttt{RegisterInvoice}. Figure~\ref{fig:invoice-source} gives the complete
Touba program. The two agents have the same scalar interface but different
responsibilities and authority. A host binding implements the tool by appending
the identifier to an isolated local register and returning it.

\begin{figure}[htbp]
\centering
\begin{ToubaCode}[title={Invoice registration: source contract},breakable=false,listing options={style=ToubaSource}]
module invoices

tool RegisterInvoice
    description "Append an invoice ID to the register."
    takes invoice as Text
    gives Text

agent InvoiceClerk
    description "Record an invoice ID and return it."
    takes invoice as Text
    gives Text
    responsibility "Register the supplied invoice ID."
    uses RegisterInvoice
    callable_by agent InvoiceDesk

agent InvoiceDesk
    takes invoice as Text
    gives Text
    responsibility "Delegate invoice registration."
    uses InvoiceClerk
    callable_by strategy ReceiveInvoice

strategy ReceiveInvoice
    takes invoice as Text
    gives Text
    uses InvoiceDesk
    do
        InvoiceDesk
            with invoice as invoice
            into registered
        give registered
\end{ToubaCode}
\caption{An executable invoice-registration boundary. Lines 3--6 declare
the external effect and its interface; lines 8--21 separate the two agents'
responsibilities and authority; lines 23--31 provide the entry strategy.
The host implementation is separate from these source contracts.}
\label{fig:invoice-source}
\end{figure}

Delegation requires agreement on both sides: the coordinator declares
\texttt{uses InvoiceClerk}; the specialist admits it through
\texttt{callable\_by}. Neither clause grants the coordinator direct access to
\texttt{RegisterInvoice}. The natural-language responsibility states the
mandate; the use and caller relations determine executable authority.
\texttt{takes} and \texttt{gives} constrain values, not the truth of a claim
that the invoice has been registered.

\begin{figure}[htbp]
\centering
\begin{tikzpicture}[
  font=\sffamily\small,
  stage/.style={draw=ToubaCodeBorder, fill=ToubaCodeBackground,
    rounded corners=4pt, align=left, inner sep=9pt,
    text width=\dimexpr\linewidth-22pt\relax},
  flow/.style={-{Stealth[length=4pt]}, draw=ToubaCodeAccent, line width=.7pt}
]
\node[stage] (source) {
  \textcolor{ToubaCodeAccent}{\textbf{1\quad Source: the specialist's boundary}}\\[3pt]
  \texttt{agent InvoiceClerk}\\
  \texttt{uses RegisterInvoice}\quad;
  \texttt{callable\_by agent InvoiceDesk}
};
\node[stage, below=11pt of source] (air) {
  \textcolor{ToubaCodeAccent}{\textbf{2\quad AIR: resolved contract fields}}\\[3pt]
  \texttt{id: agent:invoices/InvoiceClerk}\\
  \texttt{uses: [tool:invoices/RegisterInvoice]}\\
  \texttt{callable\_by[0].caller\_ref:}\\
  \hspace*{1em}\texttt{agent:invoices/InvoiceDesk}
};
\node[stage, below=11pt of air] (runtime) {
  \textcolor{ToubaCodeAccent}{\textbf{3\quad Runtime: check before execution}}\\[3pt]
  Delegation: Desk may select Clerk; Clerk admits Desk.\\
  Tool proposal: Clerk may select RegisterInvoice.\\
  The runtime supplies the active agent; the host supplies the binding.
};
\node[stage, below=11pt of runtime] (trace) {
  \textcolor{ToubaCodeAccent}{\textbf{4\quad Trace: observed identity and result fields}}\\[3pt]
  \texttt{agent\_invocation\_authorized}\\
  \hspace*{1em}\texttt{caller\_ref}: Desk; \texttt{agent\_ref}: Clerk\\[2pt]
  \texttt{agent\_capability\_result\_returned}\\
  \hspace*{1em}\texttt{agent\_ref}: Clerk; \texttt{selected\_ref}: RegisterInvoice\\
  \hspace*{1em}\texttt{result\_available: true}
};
\draw[flow] (source.south) -- (air.north);
\draw[flow] (air.south) -- (runtime.north);
\draw[flow] (runtime.south) -- (trace.north);
\end{tikzpicture}
\caption{Selected AIR and trace fields from actual compilation and execution
of Figure~\ref{fig:invoice-source}. In the lower panels, Desk, Clerk and
RegisterInvoice abbreviate the fully qualified references above. A separate
read of the local register confirms the entry \texttt{INV-042}.}
\label{fig:invoice-preservation}
\end{figure}
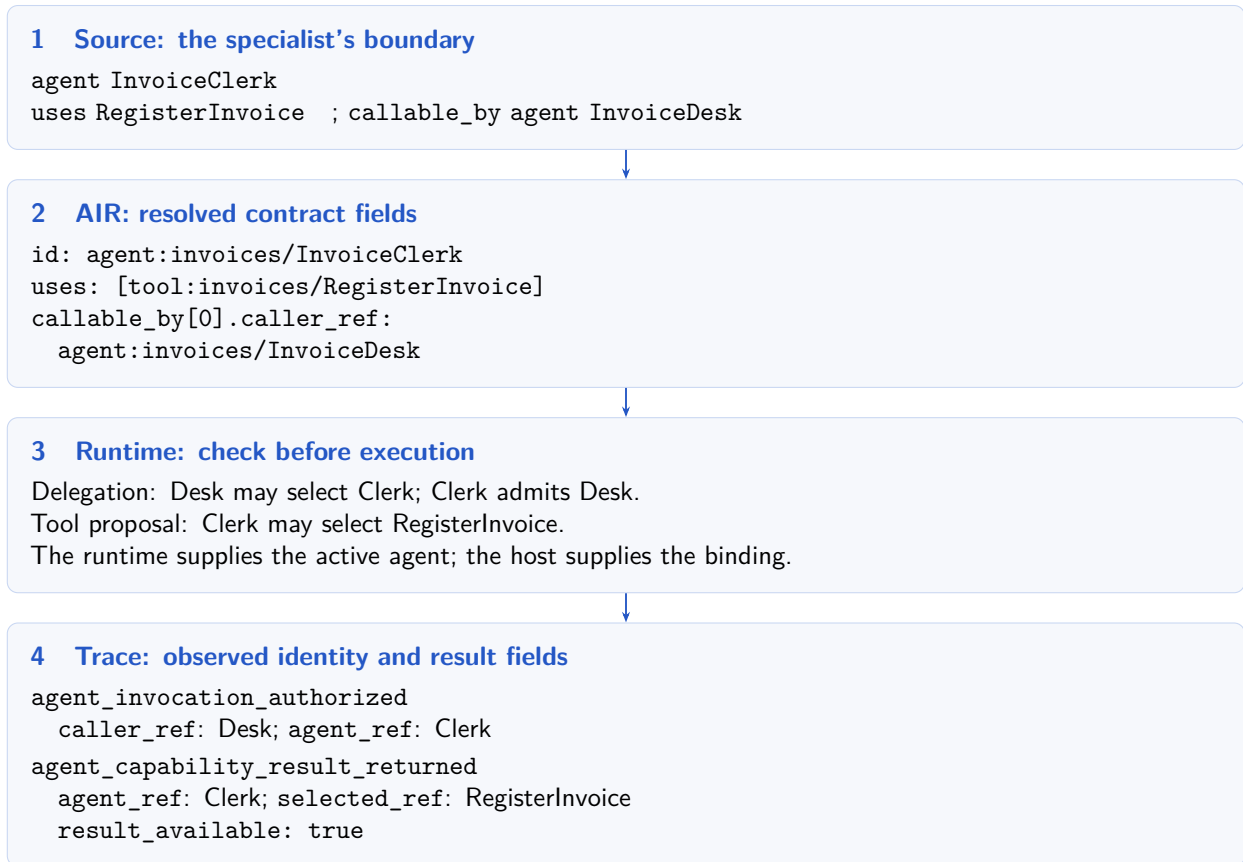

Figure~\ref{fig:invoice-preservation} follows these contracts through the
actual compiler and runtime. Fully qualified references replace local names
in AIR; the runtime checks the coordinator's selection and the specialist's
caller relation before entering the specialist. The later tool selection
belongs to the specialist. The trace projection retains these identities,
while reading the local register independently establishes the write.

This illustrates the responsibility and caller correspondence specified in
Section~\ref{sec:boundary-preservation}. Dynamic selections appear as agent
events; the explicit strategy invocation also links to a compiled node and
its source span. The returned tool observation and the independently read
register distinguish an executed effect from a selected capability.

\FloatBarrier
\subsection{When a Failed Proposal Demonstrates a Working Boundary}
\label{sec:contained-failures}

\begin{figure}[htbp]
\centering
\resizebox{\linewidth}{!}{%
\begin{tikzpicture}[
  box/.style={draw=black!20,rounded corners=3pt,fill=black!2,
    text width=4cm,minimum height=1.85cm,align=left,inner sep=7pt,
    font=\sffamily\footnotesize},
  heading/.style={text width=4cm,align=center,font=\sffamily\bfseries\small},
  arrow/.style={-{Latex[length=2mm]},draw=black!50,semithick}]
  \node[heading] at (0,0.7) {A. Delegation\\authority};
  \node[heading] at (4.8,0.7) {B. Approval\\and grants};
  \node[heading] at (9.6,0.7) {C. Reusable\\capability};
  \node[box] (ap) at (0,-0.65) {\textbf{Proposal}\\Desk directly selects the tool declared for Clerk.};
  \node[box] (bp) at (4.8,-0.65) {\textbf{Proposal}\\After human approval, Clerk selects the journal-writing skill.};
  \node[box] (cp) at (9.6,-0.65) {\textbf{Proposal}\\After one write, Clerk selects the same permitted tool again.};
  \node[box] (ag) at (0,-2.95) {\textbf{Admission check}\\Tool absent from Desk's uses. Capability refusal recorded.};
  \node[box] (bg) at (4.8,-2.95) {\textbf{Admission check}\\Required write grant absent. Resource refusal recorded.};
  \node[box] (cg) at (9.6,-2.95) {\textbf{Admission check}\\One-call quota already consumed. Dispatch refusal recorded.};
  \node[box,fill=blue!5] (ae) at (0,-5.25) {\textbf{Observed effects}\\0 handler calls.\\Register remains empty.};
  \node[box,fill=blue!5] (be) at (4.8,-5.25) {\textbf{Observed effects}\\0 process entries.\\Journal remains empty.};
  \node[box,fill=blue!5] (ce) at (9.6,-5.25) {\textbf{Observed effects}\\1 handler call in total.\\First entry retained.};
  \foreach \p/\g/\e in {ap/ag/ae,bp/bg/be,cp/cg/ce} {
    \draw[arrow] (\p.south) -- (\g.north);
    \draw[arrow] (\g.south) -- (\e.north);
  }
  \draw[black!30,dashed] (-2.2,-6.55) -- (11.8,-6.55);
  \node[box,fill=white] at (0,-7.85) {\textbf{Permitted control}\\Desk delegates to Clerk:\\1 call, 1 entry.};
  \node[box,fill=white] at (4.8,-7.85) {\textbf{Permitted control}\\Required grant supplied:\\1 process entry, 1 write.};
  \node[box,fill=white] at (9.6,-7.85) {\textbf{Permitted control}\\Quota raised to two:\\2 calls, 2 entries.};
\end{tikzpicture}%
}
\caption{Three verified governance boundaries, from proposal to observed
effects. A and C use Figure~\ref{fig:invoice-source}; B uses the compiled
journal variant with a human checkpoint and resource contracts. Each control
changes only the proposal (A), binding grant (B), or quota (C). The prevented
operation and the permitted control together establish the tested boundary.}
\label{fig:invoice-outcomes}
\end{figure}
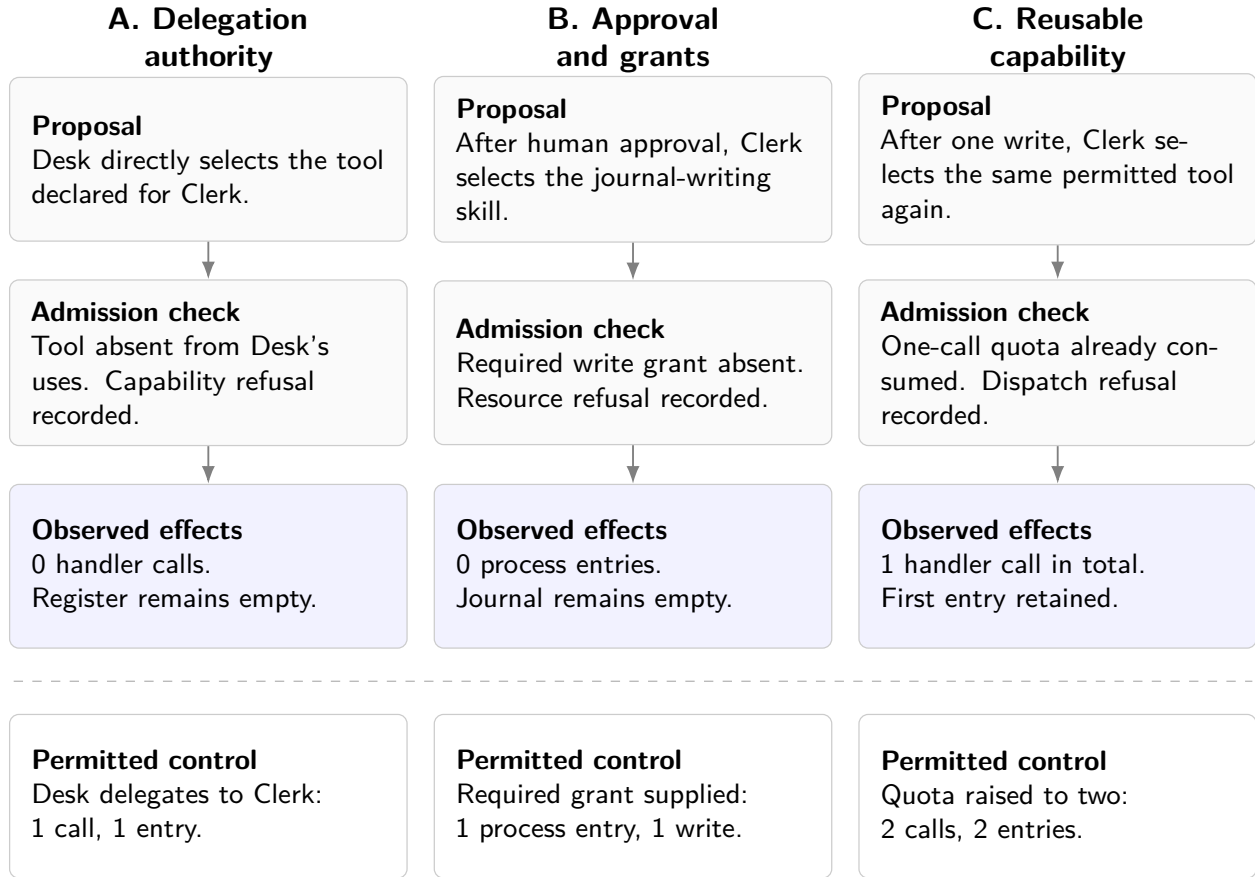

A failed run can be a successful enforcement outcome: the proposed effect
is inadmissible and the runtime prevents it. The scripted enforcement tests
in Figure~\ref{fig:invoice-outcomes} follow three such executions from proposal
to independently checked state.
The first and third use the invoice program above; the second uses a compiled
journal variant with a human checkpoint and memory/secret contracts. Each
execution starts with a fresh empty register. Provider decisions are scripted
to exercise the specified path; compilation, runtime checks, handler execution,
and file writes use the actual toolchain. The examples isolate enforcement,
rather than measuring how often a live model makes these proposals.
The panels group selected request, decision and event fields under editorial
keys; \texttt{outcome} combines terminal status with independently measured
handler counts and register contents. Tests check the displayed fields against
these executions.

\noindent\textbf{A: delegation preserves the specialist's authority.}\quad
Given \texttt{INV-042}, the coordinator proposes a direct call to
\texttt{RegisterInvoice}. Its allowed capability is \texttt{InvoiceClerk},
not the tool. The following exchange exposes the mismatch between the request's
capability list and the provider's selection, and the resulting refusal.

\begin{ToubaCode}[title={A: direct selection outside the declared authority},breakable=false]
{
  "request": {
    "agent_ref": "agent:invoices/InvoiceDesk",
    "capability_refs": ["agent:invoices/InvoiceClerk"],
    "inputs": {"invoice": "INV-042"}
  },
  "decision": {
    "action": "select_capability",
    "target_ref": "tool:invoices/RegisterInvoice",
    "arguments": {"invoice": "INV-042"}
  },
  "refusal": {
    "kind": "agent_capability_refused",
    "payload": {
      "reason": "capability_not_declared_in_uses"
    }
  },
  "outcome": {
    "status": "failure",
    "runtime_error_code": "AgentCapabilityNotAllowed",
    "handlerCalls": 0, "register": []
  }
}
\end{ToubaCode}

The handler invocation count is zero and the register remains empty.
With the same source, binding, input, and
budget, a proposal to delegate instead reaches the specialist and produces
one entry. The argument type and the tool's existence are unchanged; the
selecting agent's authority decides which path is executable.

\noindent\textbf{B: approval and resource grants work together.}\quad
In the journal variant, a mandatory checkpoint
(Section~\ref{sec:boundary-preservation}) precedes Desk, which delegates
to Clerk; Clerk selects Persist, a skill that calls Store. Store requires
write access to Journal and a secret handle, as declared in this source excerpt.

\begin{ToubaCode}[title={B: the tool's resource contract},breakable=false,listing options={style=ToubaSource,numbers=none}]
tool Store
    uses memory Journal write
    takes note as Text
    gives Text
    requires memory_write, secret_read
    secrets LEDGER_TOKEN
\end{ToubaCode}

Initially execution pauses
without invoking either the provider or the handler. A favorable decision
resumes the continuation, but a binding missing \texttt{memory\_write}
still fails. The next panel shows the approved event, the subsequent skill
selection and the resource failure. Only the binding's grants are displayed;
its memory access and secret configuration are otherwise unchanged.

\begin{ToubaCode}[title={B: approved continuation, missing write grant},breakable=false]
{
  "binding": {"grants": ["secret_read"]},
  "approval": {
    "kind": "human_approval_decided",
    "payload": {"decision": "approved"}
  },
  "decision": {
    "action": "select_capability",
    "target_ref": "skill:governed/Persist",
    "arguments": {"note": "INV-042"}
  },
  "failure": {
    "kind": "failure_detected",
    "payload": {"runtime_error_code": "MemoryGrantMissing"}
  },
  "outcome": {
    "status": "failure", "handlerCalls": 0,
    "register": []
  }
}
\end{ToubaCode}

An independent process-entry counter remains
zero and the journal is empty. Restoring that grant while keeping the other
contracts satisfied permits one write. A separate variant grants the secret
to Desk rather than Clerk: it fails with \texttt{SecretNotGrantedToAgent},
also before handler entry. Neither human approval nor an ancestor's permissions
substitutes for the active agent's resource authority.

\noindent\textbf{C: repeated use respects the declared quota.}\quad
The clerk writes the identifier once, receives the observation, and selects
the same tool again. The panel shows the runtime limit, the prior observation
returned to the provider, its second proposal and the budget refusal. The
consumed count has reached the limit; only the first write is present.

\begin{ToubaCode}[title={C: a second proposal meets an exhausted quota},breakable=false]
{
  "runLimits": {"max_tool_calls": 1},
  "prior_observation": {"value": "INV-042"},
  "decision": {
    "action": "select_capability",
    "target_ref": "tool:invoices/RegisterInvoice",
    "arguments": {"invoice": "INV-042"}
  },
  "refusal": {
    "kind": "dispatch_refused_budget_exceeded",
    "payload": {
      "limit_value": 1, "consumed_value": 1,
      "runtime_error_code": "ToolCallBudgetExceeded"
    }
  },
  "outcome": {
    "status": "failure", "handlerCalls": 1,
    "register": [{"invoice": "INV-042"}]
  }
}
\end{ToubaCode}

With a two-call limit, both writes occur.
The control demonstrates reuse under fresh occurrences, not business-level
deduplication. The first effect is neither rolled back nor erased by the
later refusal. A companion checkpoint execution verifies that a human pause
and approved resume preserve the tool counter rather than replenishing it.

\FloatBarrier

These executions instantiate different premises of
Eq.~\ref{eq:boundary-guard}: declared use in A, resource grants/scopes in B,
and \(n_t<L_t\) in C. A failed premise prevents dispatch even when the
proposal is well typed or another obligation has been satisfied. Trace order locates
the refusal, while the independent counters and registers establish what
actually ran. The authorized controls demonstrate that enforcement preserves
useful paths.

The following section evaluates this reference path through business
outcomes and artifact validation.

\section{Execution Study and Artifact Validation}
\label{sec:benchmark-evidence}

The evaluation connects the responsibility-boundary abstraction to two kinds
of executable evidence. Business cases measure whether live agents produce
the requested state. Deterministic implementation checks exercise enforcement
of declared governance contracts. The architectural comparison in the
preceding sections supplies the representation perspective.

\subsection{Business Cases and Common Conditions}

Seven cases cover warranty support, dispatch planning, purchasing approval,
incident remediation, customer-preference follow-up, invoice reconciliation
and dispute creation, and read-only price comparison. They require operations
such as pagination, repeated reads with changed arguments, data-dependent
selection, approval, and persisted writes. Incident remediation and invoice
reconciliation each use three sequential agents with typed handoffs. Touba
represents their boundaries through \texttt{takes}/\texttt{gives},
stage-scoped \texttt{uses}, \texttt{callable\_by}, and \texttt{strategy};
framework adapters use their native agent APIs with corresponding
responsibilities and operation subsets.

All systems receive the same business instructions and access to observations
through the same service operations, expressed through their respective
interfaces. Instructions,
requests, and synthetic records authored for the experiment are in English;
the task descriptions present business goals to the models. Models may reuse
authorized operations. Budgets are shared across each case's agents:
24 model-generation requests and 12 service operations, with provider retries disabled
and a 120-second request timeout. The October 2026 campaign uses seven models
from OpenAI, Anthropic, and Google. Within each model, all five methods use
the same sampling and generation limits. The GPT-4o mini and GPT-4.1 mini snapshots use
temperature zero and a 1,000-token response ceiling; the reasoning models use
low effort, omitted temperature, and model-specific output ceilings.
Appendix~\ref{app:decision-interfaces} lists the exact identifiers and profiles.
Requests use OpenAI Responses, Anthropic Messages, or Gemini's generation API
through each method's direct provider integrations. The Claude and Gemini
tool-use documentation describes their function declarations, model-generated
calls, and returned results~\cite{anthropicToolUse2026,googleGeminiFunctionCalling2026}.

Each run starts with isolated synthetic records. The verifier reloads the
persisted state and checks task-specific effects, relevant write history,
and the requested final state. The read-only price comparison instead checks
the typed final answer. This verifier is the task's \emph{oracle}: it checks
the answer or actual stored effects, rather than a claim of success in prose.
Completion requires both an oracle match
and valid termination within the shared budgets. The purchasing approver is a deterministic
simulator, and customer memory belongs to the shared service. Native review
and memory governance are exercised separately by artifact tests.
All systems face the same service constraints, although their output schemas
differ: Touba AIR carries structural types, while framework JSON schemas can
also express string, list, and numeric refinements.

Each model is evaluated with five methods: Touba SDK structured, OpenAI Agents
SDK, LangChain, LangGraph, and Touba SDK native. Five repetitions per case and
method yield 35 observations per cell. Method order rotates across cases and
repetitions; requests within a provider account are sequential.
The measured Touba SDK retains original provider messages within each
invocation and exposes typed completion in both interfaces. All five methods
share an input-token preflight over the serialized request, with the configured
output allowance reserved before generation. All 1,225 planned observations
were assessed in one campaign, including unsuccessful business outcomes.
Section~\ref{sec:validity} discusses the sampling and comparison limits.

\subsection{Observed Outcomes and Decision Interfaces}

\begin{figure}[!htbp]
\centering
\definecolor{BusinessModelBlue}{HTML}{4F96F3}
\definecolor{BusinessModelGreen}{HTML}{63BD7D}
\definecolor{BusinessModelPurple}{HTML}{9B78CB}
\begin{tikzpicture}[
  x=0.09\linewidth, y=0.077cm,
  font=\sffamily\footnotesize,
  modelA/.style={fill=BusinessModelBlue, draw=none, rounded corners=1pt},
  modelB/.style={fill=BusinessModelGreen, draw=none, rounded corners=1pt},
  modelC/.style={fill=BusinessModelPurple, draw=none, rounded corners=1pt},
  value/.style={anchor=south, inner xsep=0pt, inner ysep=2pt, text=black!85, font=\sffamily\scriptsize}
]
  \foreach \offset in {0,-80,-160} {
    \foreach \tick in {0,7,14,21,28,35} {
      \draw[black!12, line width=0.4pt, dash pattern=on 2pt off 3pt]
        (0,{\offset+\tick}) -- (10,{\offset+\tick});
      \node[anchor=east, text=black!65, inner xsep=5pt]
        at (0,{\offset+\tick}) {\tick};
    }
  }

  \begin{scope}[shift={(0,0)}]
    \node[anchor=west, font=\sffamily\small\bfseries, inner sep=0pt] at (0,54) {(a) OpenAI};
    \node[anchor=east, text=black!65, inner sep=0pt] at (10,54) {Successful runs (out of 35)};
    \fill[BusinessModelBlue] (0.2,44) circle[radius=2.8pt];
    \node[anchor=west, inner sep=0pt] at (0.45,44) {GPT-4o mini};
    \fill[BusinessModelGreen] (3.6,44) circle[radius=2.8pt];
    \node[anchor=west, inner sep=0pt] at (3.85,44) {GPT-4.1 mini};
    \fill[BusinessModelPurple] (7.2,44) circle[radius=2.8pt];
    \node[anchor=west, inner sep=0pt] at (7.45,44) {GPT-6 Astra};
    \foreach \system/\first/\second/\third/\group in {
      {Touba SDK\\structured}/21/34/35/1,
      {OpenAI Agents\\SDK}/21/35/35/3,
      {LangChain}/25/35/35/5,
      {LangGraph}/30/31/35/7,
      {Touba SDK\\native}/30/35/35/9
    } {
      \node[anchor=north, inner sep=0pt, text=black!75] at (\group,-3) {\shortstack[c]{\system}};
      \path[modelA] ({\group-0.76},0) rectangle ({\group-0.34},\first);
      \path[modelB] ({\group-0.21},0) rectangle ({\group+0.21},\second);
      \path[modelC] ({\group+0.34},0) rectangle ({\group+0.76},\third);
      \node[value] at ({\group-0.55},\first) {\first};
      \node[value] at (\group,\second) {\second};
      \node[value] at ({\group+0.55},\third) {\third};
    }
  \end{scope}

  \begin{scope}[shift={(0,-80)}]
    \node[anchor=west, font=\sffamily\small\bfseries, inner sep=0pt] at (0,54) {(b) Anthropic};
    \node[anchor=east, text=black!65, inner sep=0pt] at (10,54) {Successful runs (out of 35)};
    \fill[BusinessModelBlue] (0.2,44) circle[radius=2.8pt];
    \node[anchor=west, inner sep=0pt] at (0.45,44) {Claude Haiku 5.5};
    \fill[BusinessModelGreen] (5.2,44) circle[radius=2.8pt];
    \node[anchor=west, inner sep=0pt] at (5.45,44) {Claude Sonnet 5.5};
    \foreach \system/\first/\second/\group in {
      {Touba SDK\\structured}/31/35/1,
      {OpenAI Agents\\SDK}/34/35/3,
      {LangChain}/33/35/5,
      {LangGraph}/35/35/7,
      {Touba SDK\\native}/35/35/9
    } {
      \node[anchor=north, inner sep=0pt, text=black!75] at (\group,-3) {\shortstack[c]{\system}};
      \path[modelA] ({\group-0.68},0) rectangle ({\group-0.08},\first);
      \path[modelB] ({\group+0.08},0) rectangle ({\group+0.68},\second);
      \node[value] at ({\group-0.38},\first) {\first};
      \node[value] at ({\group+0.38},\second) {\second};
    }
  \end{scope}

  \begin{scope}[shift={(0,-160)}]
    \node[anchor=west, font=\sffamily\small\bfseries, inner sep=0pt] at (0,54) {(c) Google};
    \node[anchor=east, text=black!65, inner sep=0pt] at (10,54) {Successful runs (out of 35)};
    \fill[BusinessModelBlue] (0.2,44) circle[radius=2.8pt];
    \node[anchor=west, inner sep=0pt] at (0.45,44) {Gemini 3.8 Flash};
    \fill[BusinessModelGreen] (5.2,44) circle[radius=2.8pt];
    \node[anchor=west, inner sep=0pt] at (5.45,44) {Gemini 4 Argon};
    \foreach \system/\first/\second/\group in {
      {Touba SDK\\structured}/35/35/1,
      {OpenAI Agents\\SDK}/33/35/3,
      {LangChain}/30/35/5,
      {LangGraph}/35/35/7,
      {Touba SDK\\native}/35/35/9
    } {
      \node[anchor=north, inner sep=0pt, text=black!75] at (\group,-3) {\shortstack[c]{\system}};
      \path[modelA] ({\group-0.68},0) rectangle ({\group-0.08},\first);
      \path[modelB] ({\group+0.08},0) rectangle ({\group+0.68},\second);
      \node[value] at ({\group-0.38},\first) {\first};
      \node[value] at ({\group+0.38},\second) {\second};
    }
  \end{scope}
\end{tikzpicture}
\caption{Successful business runs: (a) OpenAI, (b) Anthropic, (c) Google.
Each bar counts completions across seven English cases repeated five times
(maximum 35). All panels share the same scale. Bar colors and left-to-right
order follow each panel's model legend. Structured and native denote Touba's
two decision interfaces. The panels cover 1,225 assessed runs;
Figure~\ref{fig:paired-interfaces} compares the Touba interfaces.}
\label{fig:english-interface-replay}
\end{figure}
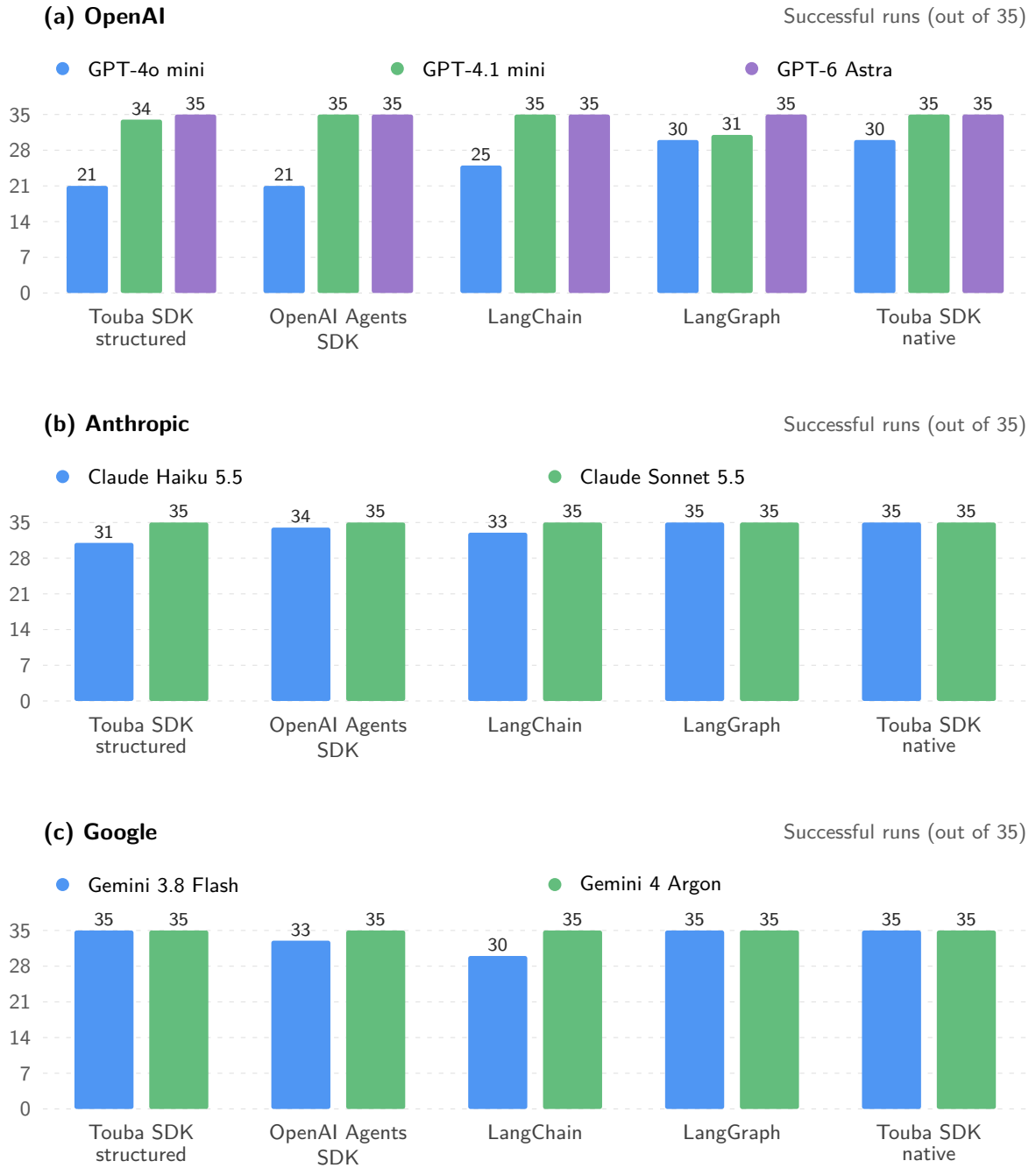

Figure~\ref{fig:english-interface-replay} reports exact business completion.
All five methods belong to the same repeated, case-paired campaign.

Native Touba matches the highest observed count for each model, including
ties with the frameworks. Its count exceeds structured Touba on GPT-4o mini,
GPT-4.1 mini and Haiku; the other four models tie. All five methods reach
the same ceiling on Astra, Sonnet and Argon.

\begin{figure}[H]
\centering
\resizebox{\linewidth}{!}{%
\begin{tikzpicture}[font=\sffamily\footnotesize]
  \node[anchor=west,font=\sffamily\bfseries\footnotesize] at (0,0.65) {Model};
  \node[align=center,font=\sffamily\bfseries\footnotesize] at (5.8,0.65) {Native minus structured};
  \node[align=center,font=\sffamily\bfseries\footnotesize] at (10.5,0.65) {Difference [95\% interval]};
  \node[align=center,font=\sffamily\bfseries\footnotesize] at (14,0.65) {Cases/run\\S / N};
  \foreach \tick in {0,20,40,60} {
    \draw[black!12] ({3.4+0.08*\tick},0.15) -- ({3.4+0.08*\tick},-5.65);
    \node[anchor=north,text=black!65] at ({3.4+0.08*\tick},-5.8) {\tick};
  }
  \draw[black!50,dashed] (3.4,0.15) -- (3.4,-5.65);
  \foreach \model/\delta/\lo/\hi/\sr/\nr/\row in {
    {GPT-4o mini}/25.7/0/57.1/{4 to 5}/6/0,
    {GPT-4.1 mini}/2.9/0/8.6/{6 to 7}/7/1,
    {GPT-6 Astra}/0/0/0/7/7/2,
    {Claude Haiku 5.5}/11.4/0/28.6/{5 to 7}/7/3,
    {Claude Sonnet 5.5}/0/0/0/7/7/4,
    {Gemini 3.8 Flash}/0/0/0/7/7/5,
    {Gemini 4 Argon}/0/0/0/7/7/6} {
    \pgfmathsetmacro{\yy}{-0.35-0.8*\row}
    \node[anchor=west] at (0,\yy) {\model};
    \draw[blue!65!black,line width=1.1pt] ({3.4+0.08*\lo},\yy) -- ({3.4+0.08*\hi},\yy);
    \draw[blue!65!black] ({3.4+0.08*\lo},{\yy-0.1}) -- ({3.4+0.08*\lo},{\yy+0.1});
    \draw[blue!65!black] ({3.4+0.08*\hi},{\yy-0.1}) -- ({3.4+0.08*\hi},{\yy+0.1});
    \fill[blue!65!black] ({3.4+0.08*\delta},\yy) circle (2.5pt);
    \node at (10.5,\yy) {\delta\ [\lo, \hi]};
    \node at (14,\yy) {\sr\ / \nr};
  }
  \node at (5.8,-6.5) {Difference in percentage points};
  \node[anchor=west,text=black!65] at (0,-7.15) {Dot: observed difference. Line: paired whole-case bootstrap interval.};
\end{tikzpicture}%
}
\caption{Native (N) minus structured (S) completion rates. Positive values
favor native mode. Intervals use 10,000 bootstrap samples of the seven whole
cases, retaining all five repetitions of each sampled case. The right column
shows the range of completed cases per repetition, out of seven.
All intervals include zero; a zero-width interval reflects a shared ceiling
in this corpus.}
\label{fig:paired-interfaces}
\end{figure}
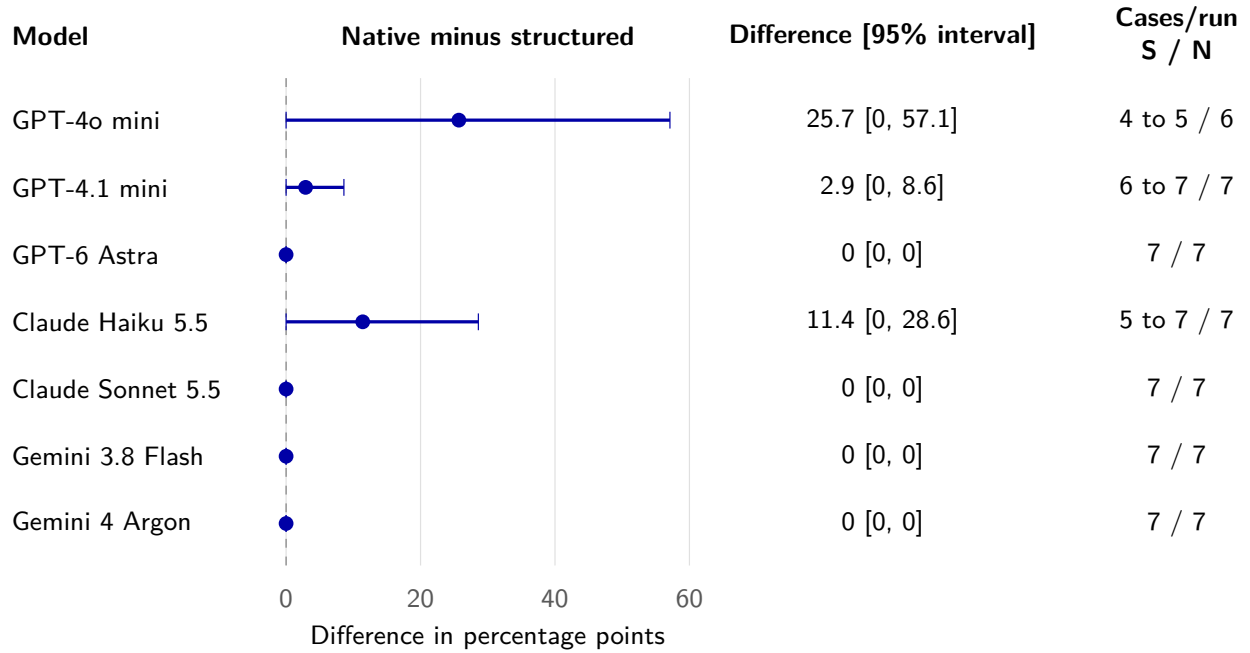

Figure~\ref{fig:paired-interfaces} summarizes the paired completion differences,
descriptive 95\% percentile intervals and repetition ranges. The whole-case
bootstrap uses the fixed seed 20261016; its sampling unit and procedure are
specified in the caption.

\FloatBarrier

The 67 unsuccessful observations comprise 31 business mismatches and 36
budget exhaustions. Table~\ref{tab:failure-breakdown} separates these outcomes
by model and method. All 36 exhaustions reached the common service-operation
limit: 16 in Agents SDK, 15 in LangChain, and five in LangGraph. They were
stopped by the shared experimental budget control. The recorded Touba traces
contain no admission refusals in this business campaign; the refusal evidence
comes from the scripted enforcement tests in
Section~\ref{sec:contained-failures}.

\begin{table}[htbp]
\centering
\small
\setlength{\tabcolsep}{4pt}
\begin{tabular}{@{}lrrrrr@{}}
\toprule
Model & Touba S & Agents SDK & LangChain & LangGraph & Touba N \\
\midrule
GPT-4o mini & 14/0 & 0/14 & 0/10 & 0/5 & 5/0 \\
GPT-4.1 mini & 1/0 & 0/0 & 0/0 & 4/0 & 0/0 \\
GPT-6 Astra & 0/0 & 0/0 & 0/0 & 0/0 & 0/0 \\
Claude Haiku 5.5 & 4/0 & 1/0 & 2/0 & 0/0 & 0/0 \\
Claude Sonnet 5.5 & 0/0 & 0/0 & 0/0 & 0/0 & 0/0 \\
Gemini 3.8 Flash & 0/0 & 0/2 & 0/5 & 0/0 & 0/0 \\
Gemini 4 Argon & 0/0 & 0/0 & 0/0 & 0/0 & 0/0 \\
\midrule
Total & 19/0 & 1/16 & 2/15 & 4/5 & 5/0 \\
\bottomrule
\end{tabular}
\caption{Unsuccessful observations, written as business mismatches / operation-budget
exhaustions. Each model--method cell contains 35 observations. S and N denote
Touba structured and native. Provider records report no output-limit truncation.}
\label{tab:failure-breakdown}
\end{table}

The five native Touba failures all occur in purchasing
with GPT-4o mini: an approval precedes an order, but their recorded values
fail the requested business conditions. Structured failures also include
incorrect remediation and preference values, and completion before a required
dispatch or follow-up write. These outcomes illustrate the distinction between
an allowed operation and a correct business decision.

Across all methods, the 7,129 recorded generation responses have normal
completion or tool-use stop indicators; none reports output-limit truncation.
For the two mini models, the largest Touba structured responses contain 93
and 161 output tokens, respectively; native maxima are 73 and 87, against
the common 1,000-token ceiling. Thus recorded cutoffs do not explain their
completion gaps. Isolating subtler effects of output allowance or response
representation would require varying those settings on the same model.

Appendix~\ref{app:decision-interfaces} gives the SDK configurations and discusses
hypotheses for representation differences.
\FloatBarrier

\subsection{Artifact Validation through Invariant Coverage}
\label{sec:artifact-validation}

\paragraph*{Language and SDK checks.}
Compilation, declaration metadata, types, imports, control flow, serialization,
and typed completion
are tested as implementation behavior. V1 construct coverage is documented
with the language and can grow in later releases.

\paragraph*{Governance checks.}
These tests follow declared contracts through source, AIR, runtime admission,
and recorded effects. They pair an inadmissible proposal with a permitted
control: authority, human review, quotas, memory scope, secrets, effect identity,
and replay remain runtime responsibilities in both SDK representations.
Figure~\ref{fig:invoice-outcomes} shows three such sequences.
A blocked operation counts as successful enforcement when the expected
guard intervenes and the forbidden effect is absent.

Table~\ref{tab:structural-evidence} describes the complementary runtime and CLI
coverage by the obligation exercised. The checks use deterministic inputs,
including refused operations and altered artifacts. They serve as software
validation of Touba's implementation; behavioral comparison uses the common
business oracle above.

\begin{table}[htbp]
\centering
\small
\begin{tabular}{@{}>{\raggedright\arraybackslash}p{0.28\linewidth}>{\raggedright\arraybackslash}p{0.66\linewidth}@{}}
\toprule
Contract area & Checked obligation \\
\midrule
Trace forensics & Reconstruct dispatch, refusal, and review pause from events. \\
Trigger identity & Accept a valid event; refuse replay and source mismatch. \\
Memory provenance & Preserve metadata and refuse raw-content or secret leakage into evidence. \\
Memory contamination & Keep injected memory content out of public traces across reads and writes. \\
Runtime budgets & Refuse excess tool calls, agent invocations, and dispatch after timeout. \\
Supply-chain integrity & Refuse a missing lock or changed handler and manifest artifacts. \\
Controlled stop & Enforce cancellation and kill switch at the tested dispatch boundaries. \\
Delegation and review & Enforce caller authority and preserve waiting continuations across review. \\
\bottomrule
\end{tabular}
\caption{Selected implementation invariants covered by deterministic runtime
and CLI tests.}
\label{tab:structural-evidence}
\end{table}

This coverage connects the formal vocabulary to inspectable implementation
behavior. For example, the memory tests check containment of injected content
in evidence, while the business oracle checks the resulting customer record.
The two checks answer complementary questions about one responsibility
boundary.

\section{Discussion and Validity}
\label{sec:validity}

\subsection{What the Evidence Establishes}

Two findings connect the formal and executable results. First, responsibility
contracts connect permission checks and responsibility across composed calls: the refused
executions identify the failed condition, the responsible context and the
effect prevented. Their authorized controls show that the same boundary
permits valid work. Second, a fixed authority boundary supports different
model-facing representations and observed completion counts.

The architectural comparison locates these contracts in programming
traditions, documented frameworks and the inspected adapters. Its evidence
concerns representation choices. Measuring an advantage over an equivalently
governed host-language implementation requires a matched comparison.

\subsection{Experimental Scope}

The main study comprises seven synthetic business cases, seven models from
three providers, five methods, and five repetitions: 1,225 assessed
observations on seven distinct tasks. The cases were visible during
development. Method order rotates but is not randomized. Shared instructions,
within-model settings, budgets, and service constraints reduce variation;
conversation structure and the constraints expressible in each schema still
vary with the adapters. Touba retains original provider messages and metadata
within bounded invocation history. Two OpenAI identifiers are dated snapshots;
the other model identifiers may evolve at their provider. Reported completion
counts every assessed attempt in the completed grid.

Exact checks of stored state establish the specified outcomes in these test
cases. New task families and randomized condition order would test
generalization. Production integrations also
introduce service failures, latency, and different data distributions.

The evaluation records retain individual outcomes, serialized model exchanges,
isolated service states, dependency versions, model profiles, and the origin
of each observation. The analysis checks the stored outcomes against their
service records before computing completion counts and paired intervals.

\subsection{Assurance Scope}

The formal results cover the chosen mediation kernel and the defined Touba
fragment in Section~\ref{sec:boundary-preservation}. Extending preservation
to the full implementation requires verification of the actual lowering,
context engine and guard evaluators for every supported construct. Compiler
rejection of effectful local functions and AIR-admission checks provide
implementation evidence for selected obligations. Durable distributed effects
and physical trace preservation require additional models.

Consistent with the founding account~\cite{lo2026foundationalPAP}, \emph{Probabilistic} identifies
systems containing probabilistic components. The kernel establishes mediation
properties for each execution path. Assigning probabilities to those paths
would require an additional probability model.

\section{Conclusion}

PAP makes the responsibility boundary an explicit programming object, linking
an agent's mandate to executable authority and attributable effects. The
comparison places this abstraction among complementary traditions in
computation, agency, orchestration and policy.

Touba's scripted enforcement tests show the practical value of that connection:
a forbidden direct call stops before handler entry, a missing grant prevents
a write after approval, and a quota blocks a repeated call while preserving
the completed effect. The traces locate each decision; independent state
checks establish its consequences. The business study complements these
examples by examining completion across model-facing representations under
fixed runtime authority.

Together, the formal model and executable evidence give a precise basis for
inspecting how responsibility is preserved through execution. The governing
principle is simple: models propose actions; declared authority governs their
effects.

Future work will examine how developers identify authority, audit effects
and modify governance using explicit responsibility boundaries.

\clearpage
\appendix
\section{Touba Decision Interfaces}
\label{app:decision-interfaces}

The following SDK configurations select the two representations introduced
in Section~\ref{sec:touba-reference}.

\subsection{Structured Decision Interface}

The default interface asks the model for one typed decision object. That object
may propose an authorized capability, request review where available, or return
a final value.

\begin{ToubaCode}[title={Structured interface: default}]
import { createAiSdkDecisionProvider } from "@touba/sdk";

export const decideAgentCapability =
  createAiSdkDecisionProvider({
    model,
    decisionInterface: "structured",
    instructions: "Use observed data and authorized capabilities.",
  });
\end{ToubaCode}

\subsection{Native Function Interface}

The opt-in native interface presents each capability already authorized for the
current step as a provider-native function with its own argument schema.

\begin{ToubaCode}[title={Native function interface: opt-in}]
import { createAiSdkDecisionProvider } from "@touba/sdk";

export const decideAgentCapability =
  createAiSdkDecisionProvider({
    model,
    decisionInterface: "native",
    instructions: "Use observed data and authorized capabilities.",
  });
\end{ToubaCode}

Both modes receive the same Touba goal, observations, and authorized catalogue,
with the runtime checks described in Section~\ref{sec:touba-reference}.

The evaluated SDK retains original provider messages, including call identifiers
and provider metadata, in a bounded history isolated to each agent invocation.

The campaign uses in-memory history without generated summaries or storage
between runs. Retention and the context sent to the model have separate bounds:
8 MiB of retained history per invocation, and a default window of 128 messages
and 262,144 bytes. A common preflight for all five methods counts the serialized
request, adds a 20\% margin and 512 tokens, and limits that input allowance to
100,000 tokens before reserving the model's output ceiling.

\subsection{Model Profiles Used in the Execution Study}

Table~\ref{tab:model-profiles} gives the request settings shared by the five
methods for each model. Output ceilings include reasoning where the provider
accounts for it within that limit. Anthropic uses adaptive thinking with low
effort and automatic native tool choice; the Touba SDK still requires a valid
typed proposal. Gemini uses low thinking; Astra uses low reasoning effort.
The temperature parameter ($T$ in the table) is omitted from requests for
these reasoning profiles.

\begin{table}[htbp]
\centering
\small
\begin{tabular}{@{}>{\raggedright\arraybackslash}p{0.57\linewidth}ll@{}}
\toprule
API model identifier & Sampling & Output limit \\
\midrule
\texttt{gpt-4o-mini-2024-07-18} & $T=0$ & 1,000 \\
\texttt{gpt-4.1-mini-2025-04-14} & $T=0$ & 1,000 \\
\texttt{gpt-6-astra} & Low effort & 4,096 \\
\texttt{claude-haiku-5-5} & Low effort & 4,096 \\
\texttt{claude-sonnet-5-5} & Low effort & 4,096 \\
\texttt{gemini-3.8-flash} & Low thinking & 8,192 \\
\texttt{gemini-4-argon} & Low thinking & 32,768 \\
\bottomrule
\end{tabular}
\caption{Model identifiers and per-request token ceilings in October 2026.
Each model uses the same profile across all five methods; profiles vary
between models.}
\label{tab:model-profiles}
\end{table}

\subsection{Interpreting Representation Differences}

Both representations use provider-supported interfaces: a JSON output schema
for structured decisions and a function-calling channel for native proposals.
This representation choice also exists outside Touba. Training familiarity is one hypothesis;
this study observes neither training data nor an isolated training effect.
The API exchange, decision schema and conversation structure provide concrete
variables for controlled tests. Table~\ref{tab:failure-breakdown} and the
recorded stop indicators distinguish business failures, operation budgets
and output-limit truncation in the present data.

A model fine-tuned on Touba structured decisions could plausibly improve its
use of that interface without changing Touba's governance boundary. Testing that proposition
requires a fixed base model, separate task families for training and evaluation,
identical prompts and budgets, paired checks of exact outcomes, and explicit
measurement of refusals and unsafe effects. This remains a direction for a
future experiment.

\FloatBarrier
\section{Mediation Kernel and Proofs}
\label{app:boundary-proofs}

This appendix specifies the abstract kernel used in
Section~\ref{sec:formal-boundary} and supplies its mathematical induction
arguments. The proofs are handwritten; the executable validation is finite
state exploration.

\subsection{State and Closed Rules}

Fix \(C=(\Gamma,\Delta)\), a run identifier, and nonnegative integer limits
\(L_p,L_t\). The state is
\[
S=(q,k,r,\chi,n_p,n_t,v,z,f,\mathit{active},\tau).
\]
Here \(q\) is the phase, \(k\) the fresh-occurrence counter, \(r\) the current
request or \(\bot\) (absence), \(\chi\) the admissible human discharge, \(n_p,n_t\)
consumed provider and tool calls, \(v\) whether a provider request is pending,
\(z\) its terminal mode, \(f\) whether the final request has been issued,
\(\mathit{active}\) the active flag, and \(\tau\) the logical audit.
The flags \(\chi,v,z,f,\mathit{active}\) are Boolean; \(k,n_p,n_t\) are
nonnegative integers. The initial state has phase \(\mathsf{idle}\), zero
counters, no request, false flags except \(\mathit{active}\), and empty audit.
Phases are \(\mathsf{idle}\), \(\mathsf{pending}\), \(\mathsf{waiting}\),
\(\mathsf{in\_flight}\), \(\mathsf{completed}\), \(\mathsf{refused}\), and
\(\mathsf{failed}\). The last three are terminal.

The following eleven rules are exhaustive. Components not mentioned are
unchanged. Every defined transition appends exactly its event to \(\tau\);
no transition rewrites \(C\). A false structural precondition gives no
transition for that input. An admission failure explicitly mentioned below
emits a refusal and enters the terminal refused phase.

\noindent\textbf{Request and reply.}\quad
REQUEST requires idle, no pending provider request, and \(f=false\).
It refuses if inactive or \(n_p\ge L_p\). Otherwise it increments \(n_p\)
before emission, sets \(v=true\), records the requested normal or terminal
mode \(z\), and sets \(f=true\) for a terminal request.
REPLY requires \(v=true\) and clears it. A malformed response, inability to
complete, or ill-typed result is refused. In terminal mode only a well-typed
final result reaches completed; a call or review is refused.
A normal-mode valid response remains data: it does not change
\(r,\chi,k,n_t\) or \(C\). The trusted engine, not the response itself, chooses
subsequent engine transitions.

\noindent\textbf{Preparation and human boundary.}\quad
OPEN requires idle, no pending provider request, and \(f=false\).
The engine supplies a candidate context and operation, assigns \(u=k\),
increments \(k\), stores \(r=(u,K,\theta,x,b,p)\), clears \(\chi\), and enters pending.
It does not certify that dispatch is allowed.
WAIT requires pending, \(\mathsf{NeedHuman}(C,r)\), and \(\chi=false\), and enters waiting.
HUMAN requires waiting and an authoritative correlated decision.
A favorable decision for the correct occurrence sets \(\chi=true\) and returns
to pending; rejection or mismatched correlation is refused.
DENY requires pending and enters refused.

\noindent\textbf{Dispatch and observation.}\quad
DISPATCH requires pending. When \(G(C,S)\) holds, it constructs \(W(C,S)\),
increments \(n_t\), emits the dispatch event, and enters in-flight; otherwise
it refuses without emitting a dispatch.
OBSERVE requires in-flight. A conforming return while active restores idle;
an unknown outcome, error, or inactive execution enters failed.
In both cases it clears \(r,\chi\) but preserves consumed counters.
An unknown result does not establish absence of an external effect.

\noindent\textbf{Completion and stop.}\quad
COMPLETE requires idle and no pending provider request. A well-typed result
while active enters completed; otherwise it refuses.
CANCEL applies to any nonterminal state, clears \(\mathit{active}\), and enters refused
unless already in-flight. An in-flight operation remains observable:
cancellation does not undo its dispatch.
FAIL enters failed from any nonterminal state.
No rule leaves a terminal state.

The engine can open successive operations of a composition without asking the
provider again. An operation parent is distinct from its fresh occurrence.
The engine also decides when finalization is appropriate; the kernel only
enforces the selected mode. Completion closes this kernel instance, while an
enclosing engine may deliver a typed sub-agent result to its parent.
Neither stack correctness nor the selection of a correct business result is
proved by these rules.

\subsection{Witness Construction}

Define
\[
W(C,S)=
(\mathit{run},r,C,(g_j(C,r))_{j\in J},\chi,\mathit{active},n_t,L_t).
\]
Here \(\mathit{run}\) is the fixed run identifier and
\((g_j(C,r))_{j\in J}\) collects the guard results. All fields refer to the
same immutable pre-state and the same trusted admission inputs, suppressed
in the notation as in Section~\ref{sec:formal-boundary}.
Let \(\mathrm{Valid}(C,S,w)\) mean exact correspondence with these fields
together with \(G(C,S)\). Contract and request references may be stored instead
of copying sensitive values into a public trace.

\noindent\textbf{Construction lemma.}\quad
If \(G(C,S)\), then \(\mathrm{Valid}(C,S,W(C,S))\).
Each required equality follows by substitution of the constructor, and
the remaining conjunction is the premise. No later observation or
provider-supplied permission is used. This lemma establishes construction
inside the model, not the truth of an unverified runtime record.

\subsection{Inductive Properties}

\noindent\textbf{Proposition 1: justified dispatch.}\quad
For every finite execution from the initial state, contracts remain \(C\)
and each dispatch has a valid pre-state witness.

\emph{Proof.}
Initially the statement is vacuous. Assume it for a prefix.
None of the eleven rules changes \(C\). DISPATCH is the only rule that can
append a dispatch event. Its positive branch requires \(G(C,S)\) and uses
\(W(C,S)\); the construction lemma gives validity. Its negative branch emits
only a refusal. The other ten rules add no dispatch, preserving the property
of the prefix. REPLY does not transfer identity or permission fields into
the operation context. The induction therefore holds for arbitrary provider
responses. \(\square\)

\noindent\textbf{Proposition 2: occurrence and human safety.}\quad
Each occurrence has at most one dispatch; a pre-dispatch denied occurrence
has none; a required human boundary needs a correlated favorable decision.

\emph{Proof.}
Only OPEN creates a request. Its strictly increasing \(k\) prevents identifier
reuse. No rule changes the request before OBSERVE clears it.
DISPATCH changes pending to in-flight, where a second DISPATCH is undefined.
A subsequent OPEN has a fresh identifier. Refusal and failure are terminal.
For the human claim, OPEN clears \(\chi\), HUMAN is its only setter, and
DISPATCH requires \(\chi\) when \(\mathsf{NeedHuman}(C,r)\). HUMAN verifies correlation and never
changes any grant. These facts are preserved by every other rule.
\(\square\)

This is at-most-once dispatch per occurrence in the crash-free model, not
exactly-once effects and not linear consumption of a capability.

\noindent\textbf{Proposition 3: budgets and terminal mode.}\quad
The counters satisfy \(0\le n_p\le L_p\) and \(0\le n_t\le L_t\).
At most one final provider request can be emitted, and its response cannot
authorize another operation.

\emph{Proof.}
Counters start at zero. Their only increments are REQUEST and DISPATCH,
which require strict inequality before adding one; other rules preserve them.
The bounds follow by induction. A terminal REQUEST sets \(f=true\);
no rule clears it. REQUEST and OPEN require \(f=false\), while REPLY
preserves the terminal mode and refuses call/review responses.
Thus neither an additional request nor an additional operation can be
created through that mode. \(\square\)

For a vector of exact counters, replace the admission test by
\(N_i+c_i(op)\le L_i\) for every affected dimension, with
\(c_i(op)\ge0\). Here \(i\) indexes a budget dimension, \(N_i\) is its
consumed amount, \(L_i\) its limit, and \(c_i(op)\) the additional cost of
operation \(op\). The same argument applies componentwise.
Accounting scope and cost must be contract-defined. Time reservations and
crash recovery require additional semantics.

\noindent\textbf{Proposition 4: ideal evidence.}\quad
The logical audit of an execution is its sequence of transition labels.

\emph{Proof.}
The initial audit is empty. Each rule appends exactly its event and preserves
the prefix. For an execution \(E\) with transition labels
\(\ell_1,\ldots,\ell_n\), induction yields
\(\tau(E)=\ell_1\cdots\ell_n\). Applying any deterministic public projection
eventwise yields the corresponding projected execution, not the redacted
values. \(\square\)

This statement assumes lossless abstract recording. Real observer failures,
crashes, and fragmented journal/checkpoint evidence are separate obligations.

\noindent\textbf{Permitted-path lemma.}\quad
If OPEN is admissible, all contract predicates hold, the execution is active,
and tool quota remains, a dispatch path exists: OPEN--DISPATCH without a human
obligation, or OPEN--WAIT--favorable HUMAN--DISPATCH with one.
A conforming observation returns to idle and permits another fresh occurrence.

\emph{Proof.}
Substitute the premises into the stated rules and
Eq.~\ref{eq:boundary-guard}. OPEN supplies pending and a fresh request;
HUMAN supplies the required discharge when necessary. DISPATCH is enabled.
OBSERVE restores idle without removing a capability from the contracts.
\(\square\)
This is existence of permitted progress, not fairness, autonomous task
success, or termination of an external component.

\FloatBarrier
\section{A Touba Fragment and Its Preservation Proof}
\label{app:preservation}

\subsection{Scope, Values, and Resolved Modules}

This appendix defines the sequential Touba realization of
Section~\ref{sec:boundary-preservation}: the fragment \(F_T\), lowering
\(L_T\), and source and graph engines using the kernel of
Appendix~\ref{app:boundary-proofs}.

A program comprises a finite, explicitly supplied, acyclic graph of local
modules. Declarations have unique names within their module and kind;
imports name a declaration and its original module. Resolution checks
existence, kind, collisions, signatures, and caller references, then retains
the transitive declaration closure. There are no aliases, wildcard imports,
re-exports, external package resolution, or runtime discovery in \(F_T\).
The canonical name \(\alpha(d)=(module,kind,name)\) is injective on declarations.
Importing \(d\) refers to the same \(\alpha(d)\), not a copy or a new authority.

Values are Boolean, Text, safe integers, finite floating-point values, finite
lists supplied as inputs or results, and finite nominal records with declared
field types. Expressions comprise literals, local or constant references,
field projection, scalar operators, and local record construction. Operators
include numeric arithmetic and comparisons, scalar equality, Boolean
connectives, Text concatenation, and scalar-to-Text conversion.
Record construction supplies exactly the declared fields; the type dependency
graph is acyclic. The fragment has no list-literal constructor, mutable fields,
object-identity equality, or host evaluation.

Fix a scalar primitive algebra \(\delta\): Int is restricted to
\([-(2^{53}-1),2^{53}-1]\), mixed arithmetic promotes Int to Float, division
produces Float, and arithmetic faults and non-finite outputs fail. Float
operations use the chosen finite floating-point semantics, not exact real
arithmetic. Both engines use this same algebra; the argument below proves
preservation of expression structure, not correctness of a hardware
arithmetic implementation. Let \(e\) denote an expression and \(\rho\) its
environment mapping in-scope names to values. The evaluator
\(\mathcal E(e,\rho)\) returns a value or an error and is defined recursively:
references look up \(\rho\), projection selects a declared field, operators
evaluate operands left to right and apply \(\delta\), and records evaluate
fields in canonical field-name order and construct a fresh typed value.
Boolean conjunction and disjunction short-circuit. Errors propagate without
evaluating a skipped operand.

The normalized statement grammar, rather than concrete surface syntax, is
\[
\begin{aligned}
 B ::={}& \epsilon \mid \mathsf{return}\ e \mid \mathsf{stop}
       \mid \mathsf{bind}\ y=e;B
       \mid y\gets\mathsf{call}\ d(\bar e);B\\
       &\mid \mathsf{if}\ e\ \{B\}\ \mathsf{else}\ \{B\};B\\
       &\mid \mathsf{for}\ y\ \mathsf{in}\ e\ \{B\};B
       \mid \mathsf{while}\ e\ \{B\};B\\
       &\mid y\gets\mathsf{human}_{m,j}(\bar e);B .
\end{aligned}
\]
Here \(B\) is a block, \(\epsilon\) the empty block, \(y\) a local binding
name, \(d\) a resolved declaration, and \(\bar e\) an ordered expression
sequence. The symbols \(::=\) and \(\mid\) introduce a grammar definition
and its alternatives; \(\gets\) binds a returned value, and \(;B\) denotes
the remaining block. Mode \(m\) is mandatory approval or an explicit review,
and \(j\) is a static checkpoint site. Each human statement has one context
expression, so its \(\bar e\) is a singleton. The name \(y\) is distinct
from the argument values \(x\) in a prepared tool operation.
Bindings are lexical; each loop visit can instantiate its
local bindings anew. Well-formedness requires resolved references, bound
local uses, argument and return types matching their declarations, Boolean
conditions, and a list-typed for-expression whose element type is assigned
to its item binding. Record fields and operators satisfy their declared
types. Functions contain no effects or human checkpoints and
call only functions. Skills and strategies may use all statement forms.
Agents carry responsibility, interfaces, uses, and admitted callers; their
responses are capability selections with arguments or typed returns.
Static calls and declared selection edges are acyclic, but loops and repeated
selections are allowed. Thus bounded call depth does not imply a bounded
number of operations. Recursion, fallback/recovery, free-form interaction,
provider-initiated review, terminal provider mode, concurrency, and crashes
are outside this fragment.

\subsection{Contracts and Lowering}

Contracts include responsibility text, signatures, source locations, static
calls, uses, callers, tool grants, memory access pairs, owner memory scopes,
and required secret handles. Memory is a logical resource accessed only
through a declared tool in this realization; its backend is supplied by the
host. Secret handles belong to contracts, not to source values.
For target tool \(\theta\), let \(M_\theta\) be its required \((memory,mode)\) pairs, \(M_b\) its host
binding's pairs, and \(M_{\mathrm{owner}}\) the effective owner's declared pairs.
Read and write are separate modes. Let \(\Sigma_\theta,\Sigma_b,\Sigma_{\mathrm{agent}}\) denote required,
binding-admitted, and active-agent-admitted secret handles.

\(L_T\) first resolves modules, copies contract fields, and maps references
by \(\alpha\). It translates each expression recursively into the same
constructor over translated references. For a statement position \(i\) in
declaration \(d\), let \(\beta_d(i)\) be its translated entry node and
\(o\) the map recovering that node's source origin:
\(o(\beta_d(i))=(d,i)\). The notation \(\beta\) denotes this family of
node maps. Bind, return, call, and human statements
become their corresponding nodes. A branch has true/false successors to its
translated arms; their normal exits join the translated continuation.
A loop has a test/header, a body edge, back edges from normal body exits, and
an exit edge to its continuation. Return and stop have no successors.
This continuation-passing construction handles nested branches and loops;
a back edge is not a new call or a new declaration. Loop-local dead bindings
may remain in graph storage but are never referenced outside their lexical
scope.

Human nodes retain mode, static site, context expressions, result binding,
and continuation. Static sites, including those in imported graphs, have
distinct qualified keys. Uses, callers, memory identities and scopes are
transported, not enlarged, by import assembly.
Host contracts are transported by the same renaming:
\(\Delta_A(\alpha(\theta))=\Delta(\theta)\), where \(\Delta_A\) is the host-contract
environment indexed by lowered names. Handler identity, grants, secret material,
and host policies themselves are unchanged; none is extracted from a
provider response.

The host-contract profile comprises the binding, grant, resource-access,
secret-lifecycle, and optional handler-identity checks defined below.

\noindent\textbf{Lemma 1 (resolution, contracts, and expressions).}\quad
Resolution and lowering preserve declaration identity, contracts, origins,
and continuations. In corresponding environments, source and lowered
expressions yield corresponding typed values or the same error.

\emph{Proof.}
Induct over a topological ordering of the module graph. Each import resolves
to one previously identified declaration; checking kind and unique names
prevents collisions. Taking its dependency closure does not add a use edge,
caller permission, memory pair, or secret grant. Induction over declarations
then preserves copied fields and references under \(\alpha\).

For expressions, literals and references agree immediately. Projection uses
the induction hypothesis for its base and the same nominal field definition.
For a strict operator, the hypotheses for its operands give the same values
or earliest error, after which the same \(\delta\) applies. For conjunction
or disjunction, equal left values select the same short-circuit case; only
the other case requires the right-operand hypothesis. Record fields agree
in the same canonical order and construct corresponding nominal values.
Induction on statements, parameterized by a translated continuation, gives
the stated graph: ordinary exits reach that continuation, branch exits join
it, loop exits reach it, and normal loop-body exits reach the header.
Terminal exits do not acquire an edge. Function purity is closed under the
acyclic function-call relation. \(\square\)

\subsection{Engines, Resources, and Human Checkpoints}

An engine frame is \((d,\kappa,\rho,\lambda,\eta,c,p,\nu)\): declaration,
continuation, live local environment, loop cursors, nearest active agent
(or none), caller, parent invocation, and fresh invocation identifier.
Cursors are local to an invocation and keyed by loop site. New calls start
with empty cursors; return restores the saved caller, including its cursors.
A for-header evaluates its list expression, tests the cursor against its
length, binds the current item and increments the cursor on entry, or resets
it on exit. A while-header evaluates its Boolean condition and increments
its visit count on entry, or resets it on exit. A fixed iteration limit is
checked at each header before taking either edge, matching this chosen
profile. Early return discards that invocation's cursors. No loop visit
consumes a capability declaration.

A bind evaluates and advances. A branch evaluates once and selects its arm.
Function and skill calls save the continuation and push the resolved frame,
retaining the active agent; typed return pops and binds the result. Agent
entry checks its caller, sets the active agent to the callee, and uses
normal REQUEST/REPLY to obtain response data. A selection checks uses and
argument types; an agent target also checks callers. A typed agent return
restores its parent. Invalid data fails without a callee dispatch.
Provider-supplied owners, callers, bindings, and scopes never enter frames.

Let \(H\) be host resource state and \(\omega\) trusted host input at an
admission step, including time and secret availability. Contracts \(\Delta\)
fix the resolver and lifecycle policy, not a claim that a secret never
expires. A tool observation follows a host relation
\(\mathcal{H}_\Delta(H,\theta,b,x,\omega)\) returning a value or error and new
resource state. This relation is identical under renaming in both engines;
corresponding executions use the same external outcomes. It may be
nondeterministic. No correctness of stored business data is assumed or proved.

The guards now have the following concrete interpretations.
Context checks root-linked frames, saved continuations, and the active agent.
Use checks dynamic selections against their selector and static calls
against their origin statement. Caller checks each entered agent's admitted
caller. Type recursively checks the target's scalar/list/record signature.
Binding requires the host-resolved usable handler; grant requires every
declared capability. Integrity compares required and verified host identities.
Writing the admission input explicitly, the resource guard is
\[
\begin{aligned}
g_{\mathsf{scope}}(C,r;\omega)={}&[M_\theta\subseteq M_b\cap M_{\mathrm{owner}}]
 \land[\Sigma_\theta\subseteq\Sigma_b]\\
 &{}\land[\eta=\bot\ \lor\Sigma_\theta\subseteq\Sigma_{\mathrm{agent}}]
 \land\bigwedge_{s\in\Sigma_\theta}\mathsf{Live}_{\Delta}(s,\omega).
\end{aligned}
\]
Here \(\subseteq\) means set inclusion, \(\cap\) intersection, and
\(\bot\) no active agent. For each required secret handle \(s\),
\(\mathsf{Live}_{\Delta}(s,\omega)\) is true exactly when the trusted
resolver can supply it and its lifecycle policy admits its use at the time
represented by \(\omega\); missing, inactive, or expired handles fail this
test. The semicolon makes visible the host-input parameter suppressed in
\(g_j(C,r)\): it is shared by \(G\) and \(W\) for the same dispatch,
and does not change the fixed contracts \(C\).
In words, required memory modes must be admitted by both binding and owner;
required secrets must be admitted by the binding and any active agent,
and each required handle must be live. An empty conjunction is true.
The effective owner is \(\eta\) when an agent is active, and otherwise the
structural declaration issuing the tool call. Its scope is not the union of all ancestor scopes.
Memory grants and secret-read capability are checked separately by the
grant guard. All missing required facts are false.
Secret material stays in trusted host state and is delivered only through
the selected authorized handler channel in this model. A common boundary
filter may reject payloads containing known secret values; this is not a
proof against transformed disclosure or a malicious authorized handler.

A tool call prepares its request from the frame, the current trusted host
input, and the resolved binding, then follows OPEN--DISPATCH--OBSERVE.
Observation validates the result and resumes the saved continuation.
Root return uses COMPLETE, not child return. Failures and cancellation use
the kernel rules. Only mediated tool dispatches can change \(H\).

For source human statements, the engine has a waiting record
\[
Q=(j,m,\bar e,\rho,y,\kappa,\sigma),
\]
The components are the checkpoint site, mode, context expression sequence,
saved environment, decision-result binding, continuation, and saved engine
state \(\sigma\), respectively.
On reaching the statement, context expressions and their environment are
retained and the continuation is saved before advancing. Like Touba's context
references, this record identifies context without requiring its eager
evaluation or presenting it as authenticated consent. An absent decision leaves the engine
waiting; no ordinary engine step, provider request, or dispatch is enabled.
A host decision for another site leaves this wait unchanged. A supplied
decision at \(j\) is consumed when used. Approval binds it and releases the
saved continuation; rejection,
expiration, or cancellation stops it; invalid decision data fails.
Resumption restores live values, cursors, identities and counters, rather
than re-executing the prefix. Host policies may change liveness inputs during
the wait, and the next tool admission checks them again.

The host decision map is keyed by the static site. Each use removes its
entry, so a later visit waits for a newly supplied decision. This establishes
single consumption per supplied program decision. Authentication of the human
and binding consent to exact future tool arguments are separate contracts.
For this realization \(\mathsf{NeedHuman}(C,r)=false\) at tool preparation;
Section~\ref{sec:boundary-preservation} explains the two approval contracts.

The source engine can retain its waiting state in memory. Correspondence
with Touba's exported checkpoint mechanism is limited to supported entry-graph
sites; a deeper saved call stack is not assumed serializable. The theorem
below concerns crash-free engine suspension with intact state, not persistence,
authentication of approvers, or recovery of external effects.

\subsection{Simulation and Evidence}

Let \(R\) relate source and graph engines when frames correspond under
\(\alpha,\beta,o\), live values and loop cursors correspond, saved
continuations identify the same suffix, and fresh counters, host state,
pending responses, waiting records, supplied decision maps, kernel states, and audits agree under
renaming. Extra unreachable graph slots are ignored. At an engine step,
both use the same provider response, host outcome, clock input, or human
decision. They start from the same typed inputs and fixed contracts.

\noindent\textbf{Lemma 2 (backward step simulation).}\quad
From \(R\)-related states, every graph-engine move has a source match, up to
administrative stuttering, with related successors and the same visible
dispatch or checkpoint label.

\emph{Proof.}
The following cases exhaust the statement and engine rules.
Bind, return, and branch use Lemma~1 for values, errors, and continuation
targets. For a for-header, equal lists and cursors give the same item,
increment, exit, or limit failure. For a while-header, equal conditions and
counts do likewise. Back edges correspond to re-entering the saved loop
continuation. No termination assumption is needed. Calls push fresh
corresponding frames, preserving the parent cursors; return discards callee
cursors and restores the parent even after an early loop return.

Agent entry and selection test the same use/caller relations by Lemma~1;
both install the callee as active agent and restore the same parent on
return. REQUEST/REPLY use equal kernel premises and counters. Neither copies
identity fields from the response. Static and dynamic tool preparation
therefore produce corresponding contexts and origins.

For each tool, recursive value membership gives equal type guards.
Transported host contracts give equal binding, grant, and integrity guards.
Equal active owners and the copied memory pairs yield equal \(M_\theta,M_b,M_{\mathrm{owner}}\).
Required and admitted handle sets agree; coupled \(\omega\) gives equal
liveness tests, including expiry or missing resolution. Thus resource guards
agree by their explicit definitions, not by presuming source authorization.
The same optional boundary filter rejects the same payload.
OPEN, dispatch or refusal, and OBSERVE follow with equal premises;
the same \(\mathcal H_\Delta\) outcome gives corresponding values and host
state. A failed guard or observation yields the same failure.

At a human node, Lemma~1 gives corresponding context expressions and environments, mode, qualified site,
and saved continuation. Both wait, consume the same supplied site decision, stop,
or fail. Consumption removes the matching entry from both decision maps. Waiting has no ordinary successors. Approval restores matching
cursors and counters without traversing any prior effect; it modifies
neither uses, callers, scopes, nor grants. This remains true inside a loop.
CANCEL, FAIL, and root COMPLETE have equal premises. Administrative graph
navigation has no dispatch label and reaches the source continuation by
Lemma~1. Every case preserves \(R\). \(\square\)

\emph{Proof of Theorem~1.}
Initial states are related. Induction over the finite graph execution using
Lemma~2 constructs a matching source execution and preserves \(R\).
For every dispatch, Proposition~1 of Appendix~\ref{app:boundary-proofs}
supplies the valid kernel witness. Its arguments, context, contracts,
resource decisions and pre-dispatch counters agree under \(R\).
The public evidence projection retains run and occurrence identity, origin,
tool, structural owner, active-agent and parent references, and contract
references, while omitting secret material and argument values.
Proposition~4 and equality of matched dispatch labels give equal projected
dispatch sequences after inverse renaming. The human-node case additionally
gives checkpoint continuation preservation. \(\square\)

\PaperPrintDeclarations


\begin{thebibliography}{61}
\providecommand{\natexlab}[1]{#1}
\providecommand{\url}[1]{\texttt{#1}}
\expandafter\ifx\csname urlstyle\endcsname\relax
  \providecommand{\doi}[1]{doi: #1}\else
  \providecommand{\doi}{doi: \begingroup \urlstyle{rm}\Url}\fi

\bibitem[{Anthropic}(2026)]{anthropicToolUse2026}
{Anthropic}.
\newblock Tool use with {Claude}, 2026.
\newblock URL
  \url{https://platform.claude.com/docs/en/agents-and-tools/tool-use/overview}.

\bibitem[Backus(1978)]{backus1978vonNeumann}
John Backus.
\newblock Can programming be liberated from the {von Neumann} style? a
  functional style and its algebra of programs.
\newblock \emph{Communications of the ACM}, 21\penalty0 (8):\penalty0 613--641,
  1978.
\newblock \doi{10.1145/359576.359579}.

\bibitem[Black(2013)]{black2013oopHistory}
Andrew~P. Black.
\newblock Object-oriented programming: Some history, and challenges for the
  next fifty years.
\newblock \emph{Information and Computation}, 231:\penalty0 3--20, 2013.
\newblock \doi{10.1016/j.ic.2013.08.002}.

\bibitem[Cutler et~al.(2024)Cutler, Disselkoen, Eline, He, Headley, Hicks,
  Hietala, Ioannidis, Kastner, Mamat, McAdams, McCutchen, Rungta, Torlak, and
  Wells]{cedar2024authorizationLanguage}
Joseph~W. Cutler, Craig Disselkoen, Aaron Eline, Shaobo He, Kyle Headley,
  Michael Hicks, Kesha Hietala, Eleftherios Ioannidis, John Kastner, Anwar
  Mamat, Darin McAdams, Matt McCutchen, Neha Rungta, Emina Torlak, and
  Andrew~M. Wells.
\newblock {Cedar}: A new language for expressive, fast, safe, and analyzable
  authorization.
\newblock \emph{Proceedings of the ACM on Programming Languages}, 8\penalty0
  (OOPSLA1):\penalty0 118:1--118:28, 2024.
\newblock \doi{10.1145/3649835}.
\newblock URL
  \url{https://cdn.amazon.science/96/a8/1b427993481cbdf0ef2c8ca6db85/cedar-a-new-language-for-expressive-fast-safe-and-analyzable-authorization.pdf}.

\bibitem[Dastani(2008)]{dastani2008twoAPL}
Mehdi Dastani.
\newblock {2APL}: A practical agent programming language.
\newblock \emph{Autonomous Agents and Multi-Agent Systems}, 16\penalty0
  (3):\penalty0 214--248, 2008.
\newblock \doi{10.1007/s10458-008-9036-y}.
\newblock URL
  \url{https://link.springer.com/content/pdf/10.1007/s10458-008-9036-y.pdf}.

\bibitem[Debenedetti et~al.(2024)Debenedetti, Zhang, Balunovi{\'c},
  Beurer-Kellner, Fischer, and Tram{\`e}r]{agentdojo2024benchmark}
Edoardo Debenedetti, Jie Zhang, Mislav Balunovi{\'c}, Luca Beurer-Kellner, Marc
  Fischer, and Florian Tram{\`e}r.
\newblock {AgentDojo}: A dynamic environment to evaluate prompt injection
  attacks and defenses for {LLM} agents.
\newblock In \emph{Advances in Neural Information Processing Systems},
  volume~37, 2024.
\newblock \doi{10.52202/079017-2636}.
\newblock URL
  \url{https://proceedings.neurips.cc/paper_files/paper/2024/file/97091a5177d8dc64b1da8bf3e1f6fb54-Paper-Datasets_and_Benchmarks_Track.pdf}.
\newblock Datasets and Benchmarks Track.

\bibitem[Debenedetti et~al.(2025)Debenedetti, Shumailov, Fan, Hayes, Carlini,
  Fabian, Kern, Shi, Terzis, and
  Tram{\`e}r]{camel2025defeatingPromptInjections}
Edoardo Debenedetti, Ilia Shumailov, Tianqi Fan, Jamie Hayes, Nicholas Carlini,
  Daniel Fabian, Christoph Kern, Chongyang Shi, Andreas Terzis, and Florian
  Tram{\`e}r.
\newblock Defeating prompt injections by design.
\newblock \emph{arXiv}, 2025.
\newblock \doi{10.48550/arXiv.2503.18813}.
\newblock CaMeL; arXiv v2, 24 June 2025.

\bibitem[Dijkstra(1968)]{dijkstra1968goto}
Edsger~W. Dijkstra.
\newblock Go to statement considered harmful.
\newblock \emph{Communications of the ACM}, 11\penalty0 (3):\penalty0 147--148,
  1968.
\newblock \doi{10.1145/362929.362947}.

\bibitem[{European Parliament and Council of the European
  Union}(2024)]{eu2024aiAct}
{European Parliament and Council of the European Union}.
\newblock Regulation ({EU}) 2024/1689 laying down harmonised rules on
  artificial intelligence, 2024.
\newblock URL \url{https://eur-lex.europa.eu/eli/reg/2024/1689/oj}.
\newblock Official Journal of the European Union, 12 July 2024.

\bibitem[{Executive Office of the President, Office of Management and
  Budget}(2024)]{omb2024m2410}
{Executive Office of the President, Office of Management and Budget}.
\newblock Advancing governance, innovation, and risk management for agency use
  of artificial intelligence, 2024.
\newblock URL
  \url{https://bidenwhitehouse.archives.gov/wp-content/uploads/2024/03/M-24-10-Advancing-Governance-Innovation-and-Risk-Management-for-Agency-Use-of-Artificial-Intelligence.pdf}.
\newblock Memorandum M-24-10, 28 March 2024.

\bibitem[{Executive Office of the President, Office of Management and
  Budget}(2025)]{omb2025m2521}
{Executive Office of the President, Office of Management and Budget}.
\newblock Accelerating federal use of {AI} through innovation, governance, and
  public trust, 2025.
\newblock URL
  \url{https://www.whitehouse.gov/wp-content/uploads/2025/02/M-25-21-Accelerating-Federal-Use-of-AI-through-Innovation-Governance-and-Public-Trust.pdf}.
\newblock Memorandum M-25-21, 3 April 2025; rescinds and replaces M-24-10.

\bibitem[{Google}(2026{\natexlab{a}})]{googleAdkDocs2026}
{Google}.
\newblock Agent development kit documentation, 2026{\natexlab{a}}.
\newblock URL \url{https://adk.dev/}.

\bibitem[{Google}(2026{\natexlab{b}})]{googleGeminiFunctionCalling2026}
{Google}.
\newblock Function calling with the {Gemini API}, 2026{\natexlab{b}}.
\newblock URL \url{https://ai.google.dev/gemini-api/docs/function-calling}.

\bibitem[Hardy(1988)]{hardy1988confusedDeputy}
Norm Hardy.
\newblock The confused deputy (or why capabilities might have been invented).
\newblock \emph{ACM SIGOPS Operating Systems Review}, 22\penalty0 (4):\penalty0
  36--38, 1988.
\newblock \doi{10.1145/54289.871709}.
\newblock URL
  \url{https://www.cs.utexas.edu/~witchel/S25-380L/papers/hardy88confused.pdf}.

\bibitem[Hewitt et~al.(1973)Hewitt, Bishop, and Steiger]{hewitt1973actors}
Carl Hewitt, Peter Bishop, and Richard Steiger.
\newblock A universal modular actor formalism for artificial intelligence.
\newblock In \emph{Proceedings of the 3rd International Joint Conference on
  Artificial Intelligence}, pages 235--245, 1973.
\newblock URL \url{https://www.ijcai.org/Proceedings/73/Papers/027B.pdf}.
\newblock IJCAI-73 proceedings.

\bibitem[Hofer et~al.(2026)Hofer, Debenedetti, and
  Tram{\`e}r]{hofer2026automatedPromptInjection}
David Hofer, Edoardo Debenedetti, and Florian Tram{\`e}r.
\newblock Assessing automated prompt injection attacks in agentic environments.
\newblock \emph{arXiv}, 2026.
\newblock \doi{10.48550/arXiv.2606.10525}.

\bibitem[{in-toto}(2026)]{inToto2026Framework}
{in-toto}.
\newblock A framework to secure the integrity of software supply chains, 2026.
\newblock URL \url{https://in-toto.io/}.

\bibitem[Karpas et~al.(2022)Karpas, Abend, Belinkov, Lenz, Lieber, Ratner,
  Shoham, Bata, Levine, Leyton-Brown, Muhlgay, Rozen, Schwartz, Shachaf,
  Shalev-Shwartz, Shashua, and Tenenholtz]{karpas2022mrklSystems}
Ehud Karpas, Omri Abend, Yonatan Belinkov, Barak Lenz, Opher Lieber, Nir
  Ratner, Yoav Shoham, Hofit Bata, Yoav Levine, Kevin Leyton-Brown, Dor
  Muhlgay, Noam Rozen, Erez Schwartz, Gal Shachaf, Shai Shalev-Shwartz, Amnon
  Shashua, and Moshe Tenenholtz.
\newblock {MRKL} systems: A modular, neuro-symbolic architecture that combines
  large language models, external knowledge sources and discrete reasoning.
\newblock \emph{arXiv}, 2022.
\newblock \doi{10.48550/arXiv.2205.00445}.

\bibitem[{LangChain}(2026{\natexlab{a}})]{langchainJsReview2026}
{LangChain}.
\newblock {LangChain JavaScript}: Human-in-the-loop, 2026{\natexlab{a}}.
\newblock URL
  \url{https://docs.langchain.com/oss/javascript/langchain/human-in-the-loop}.

\bibitem[{LangChain}(2026{\natexlab{b}})]{langgraphJsInterrupts2026}
{LangChain}.
\newblock {LangGraph JavaScript}: Interrupts, 2026{\natexlab{b}}.
\newblock URL
  \url{https://docs.langchain.com/oss/javascript/langgraph/interrupts}.

\bibitem[{LangChain}(2026{\natexlab{c}})]{langgraphPersistence2026}
{LangChain}.
\newblock {LangGraph} persistence, 2026{\natexlab{c}}.
\newblock URL
  \url{https://docs.langchain.com/oss/python/langgraph/persistence}.

\bibitem[Li et~al.(2026)Li, Wen, Shi, Zhang, Vorobeychik, and
  Xiao]{agentdyn2026benchmark}
Hao Li, Ruoyao Wen, Shanghao Shi, Ning Zhang, Yevgeniy Vorobeychik, and Chaowei
  Xiao.
\newblock {AgentDyn}: Are your agent security defenses deployable in real-world
  dynamic environments?
\newblock \emph{arXiv}, 2026.
\newblock \doi{10.48550/arXiv.2602.03117}.
\newblock URL \url{https://arxiv.org/abs/2602.03117v3}.
\newblock arXiv v3, 7 May 2026.

\bibitem[Lo et~al.(2026)Lo, Alaoui~Ismaili, and
  Dessureault]{lo2026foundationalPAP}
Khadim Lo, Soukaina Alaoui~Ismaili, and Jean-S{\'e}bastien Dessureault.
\newblock Probabilistic agentic programming: A foundational paradigm for
  governed autonomy.
\newblock Manuscript submitted to AI and Ethics, 2026.

\bibitem[Lucassen and Gifford(1988)]{lucassen1988effects}
John~M. Lucassen and David~K. Gifford.
\newblock Polymorphic effect systems.
\newblock In \emph{Proceedings of the 15th ACM SIGPLAN-SIGACT Symposium on
  Principles of Programming Languages}, pages 47--57, 1988.
\newblock \doi{10.1145/73560.73564}.
\newblock URL \url{https://cs.ioc.ee/ewscs/2010/mycroft/lucassen-popl88.pdf}.

\bibitem[Lynch and Vaandrager(1995)]{lynch1995simulations}
Nancy Lynch and Frits Vaandrager.
\newblock Forward and backward simulations: {I}. untimed systems.
\newblock \emph{Information and Computation}, 121\penalty0 (2):\penalty0
  214--233, 1995.
\newblock \doi{10.1006/inco.1995.1134}.
\newblock URL \url{https://ir.cwi.nl/pub/1393/1393D.pdf}.

\bibitem[Miller(2006)]{miller2006robustComposition}
Mark~Samuel Miller.
\newblock \emph{Robust Composition: Towards a Unified Approach to Access
  Control and Concurrency Control}.
\newblock PhD thesis, Johns Hopkins University, 2006.
\newblock URL \url{https://jscholarship.library.jhu.edu/handle/1774.2/873}.

\bibitem[{Model Context Protocol}(2025)]{anthropicModelContextProtocol2025}
{Model Context Protocol}.
\newblock Model context protocol specification, 2025.
\newblock URL \url{https://modelcontextprotocol.io/specification/2025-11-25}.
\newblock Version 2025-11-25.

\bibitem[{National Institute of Standards and Technology}(2023)]{nist2023aiRmf}
{National Institute of Standards and Technology}.
\newblock Artificial intelligence risk management framework ({AI RMF} 1.0).
\newblock {NIST AI} 100-1, National Institute of Standards and Technology,
  2023.
\newblock URL \url{https://doi.org/10.6028/NIST.AI.100-1}.
\newblock Published 26 January 2023.

\bibitem[{National Institute of Standards and
  Technology}(2024)]{nist2024genAiProfile}
{National Institute of Standards and Technology}.
\newblock Artificial intelligence risk management framework: Generative
  artificial intelligence profile.
\newblock {NIST AI} 600-1, National Institute of Standards and Technology,
  2024.
\newblock URL \url{https://doi.org/10.6028/NIST.AI.600-1}.
\newblock Published 26 July 2024.

\bibitem[Necula(1997)]{necula1997proofCarryingCode}
George~C. Necula.
\newblock Proof-carrying code.
\newblock In \emph{Proceedings of the 24th ACM SIGPLAN-SIGACT Symposium on
  Principles of Programming Languages}, pages 106--119, 1997.
\newblock \doi{10.1145/263699.263712}.
\newblock URL
  \url{https://homes.cs.washington.edu/%7Emernst/teaching/6.893/readings/necula-popl97.pdf}.

\bibitem[{OASIS eXtensible Access Control Markup Language Technical
  Committee}(2013)]{oasis2013xacml}
{OASIS eXtensible Access Control Markup Language Technical Committee}.
\newblock {eXtensible Access Control Markup Language} ({XACML}) version 3.0,
  2013.
\newblock URL
  \url{https://docs.oasis-open.org/xacml/3.0/xacml-3.0-core-spec-os-en.html}.
\newblock OASIS Standard, 22 January 2013.

\bibitem[{OASIS Web Services Business Process Execution Language Technical
  Committee}(2007)]{oasis2007wsbpel}
{OASIS Web Services Business Process Execution Language Technical Committee}.
\newblock Web services business process execution language version 2.0, 2007.
\newblock URL
  \url{https://docs.oasis-open.org/wsbpel/2.0/OS/wsbpel-v2.0-OS.html}.
\newblock OASIS Standard, 11 April 2007.

\bibitem[{Office of the Superintendent of Financial
  Institutions}(2025)]{osfi2025e23}
{Office of the Superintendent of Financial Institutions}.
\newblock Guideline {E-23}: Model risk management, 2025.
\newblock URL
  \url{https://www.osfi-bsif.gc.ca/en/guidance/guidance-library/guideline-e-23-model-risk-management-2027}.
\newblock Published 11 September 2025; effective 1 May 2027.

\bibitem[{Open Policy Agent}(2026)]{opa2026docs}
{Open Policy Agent}.
\newblock Open policy agent documentation, 2026.
\newblock URL \url{https://www.openpolicyagent.org/docs}.

\bibitem[{OpenAI}(2026{\natexlab{a}})]{openaiAgentsSdk2026}
{OpenAI}.
\newblock {OpenAI Agents SDK} documentation, 2026{\natexlab{a}}.
\newblock URL \url{https://developers.openai.com/api/docs/guides/agents}.

\bibitem[{OpenAI}(2026{\natexlab{b}})]{openaiSdkReview2026}
{OpenAI}.
\newblock Agents {SDK}: Guardrails and human review, 2026{\natexlab{b}}.
\newblock URL
  \url{https://developers.openai.com/api/docs/guides/agents/guardrails-approvals}.

\bibitem[{OpenAI}(2026{\natexlab{c}})]{openaiSdkTracing2026}
{OpenAI}.
\newblock Agents {SDK}: Integrations and observability, 2026{\natexlab{c}}.
\newblock URL
  \url{https://developers.openai.com/api/docs/guides/agents/integrations-observability}.

\bibitem[{OpenTelemetry}(2026)]{opentelemetryGenAi2026}
{OpenTelemetry}.
\newblock Semantic conventions for generative {AI} systems, 2026.
\newblock URL
  \url{https://github.com/open-telemetry/semantic-conventions-genai/tree/main/docs}.

\bibitem[{OWASP GenAI Security
  Project}(2025{\natexlab{a}})]{owasp2025agenticTop10}
{OWASP GenAI Security Project}.
\newblock {OWASP GenAI} security project releases top 10 risks and mitigations
  for agentic {AI} security, 2025{\natexlab{a}}.
\newblock URL
  \url{https://genai.owasp.org/2025/12/09/owasp-genai-security-project-releases-top-10-risks-and-mitigations-for-agentic-ai-security/}.
\newblock Announcement dated 9 December 2025; release dated 10 December 2025.

\bibitem[{OWASP GenAI Security
  Project}(2025{\natexlab{b}})]{owaspAgenticAiThreatsMitigations}
{OWASP GenAI Security Project}.
\newblock Agentic {AI}: Threats and mitigations, 2025{\natexlab{b}}.
\newblock URL
  \url{https://genai.owasp.org/resource/agentic-ai-threats-and-mitigations/}.
\newblock Agentic Security Initiative, 17 February 2025.

\bibitem[Parnas(1972)]{parnas1972criteria}
David~L. Parnas.
\newblock On the criteria to be used in decomposing systems into modules.
\newblock \emph{Communications of the ACM}, 15\penalty0 (12):\penalty0
  1053--1058, 1972.
\newblock \doi{10.1145/361598.361623}.

\bibitem[{Python Software Foundation}(n.d.)]{pythonTyping2026}
{Python Software Foundation}.
\newblock typing: Support for type hints.
\newblock Python documentation, n.d.
\newblock URL \url{https://docs.python.org/3/library/typing.html}.

\bibitem[Rao and Georgeff(1995)]{raoGeorgeff1995bdiAgents}
Anand~S. Rao and Michael~P. Georgeff.
\newblock {BDI} agents: From theory to practice.
\newblock In \emph{Proceedings of the First International Conference on
  Multi-Agent Systems}, pages 312--319, 1995.
\newblock URL \url{https://cdn.aaai.org/ICMAS/1995/ICMAS95-042.pdf}.
\newblock ICMAS-95, San Francisco.

\bibitem[{Restate}(2026)]{restateDurableExecution2026}
{Restate}.
\newblock Key concepts: Durable execution, 2026.
\newblock URL
  \url{https://docs.restate.dev/foundations/key-concepts#durable-execution}.

\bibitem[Sabelfeld and Myers(2003)]{sabelfeld2003informationFlow}
Andrei Sabelfeld and Andrew~C. Myers.
\newblock Language-based information-flow security.
\newblock \emph{IEEE Journal on Selected Areas in Communications}, 21\penalty0
  (1):\penalty0 5--19, 2003.
\newblock \doi{10.1109/JSAC.2002.806121}.
\newblock URL \url{https://www.cs.cornell.edu/andru/papers/jsac/sm-jsac03.pdf}.

\bibitem[Saltzer and Schroeder(1975)]{saltzer1975protection}
Jerome~H. Saltzer and Michael~D. Schroeder.
\newblock The protection of information in computer systems.
\newblock \emph{Proceedings of the IEEE}, 63\penalty0 (9):\penalty0 1278--1308,
  1975.
\newblock \doi{10.1109/PROC.1975.9939}.
\newblock URL
  \url{https://web.mit.edu/Saltzer/www/publications/protection/index.html}.

\bibitem[Schick et~al.(2023)Schick, Dwivedi-Yu, Dess{\`i}, Raileanu, Lomeli,
  Hambro, Zettlemoyer, Cancedda, and Scialom]{schick2023toolformer}
Timo Schick, Jane Dwivedi-Yu, Roberto Dess{\`i}, Roberta Raileanu, Maria
  Lomeli, Eric Hambro, Luke Zettlemoyer, Nicola Cancedda, and Thomas Scialom.
\newblock Toolformer: Language models can teach themselves to use tools.
\newblock In \emph{Advances in Neural Information Processing Systems 36}, 2023.
\newblock \doi{10.52202/075280-2997}.
\newblock URL \url{https://openreview.net/pdf?id=Yacmpz84TH}.
\newblock NeurIPS 2023.

\bibitem[Schneider(2000)]{schneider2000enforceable}
Fred~B. Schneider.
\newblock Enforceable security policies.
\newblock \emph{ACM Transactions on Information and System Security},
  3\penalty0 (1):\penalty0 30--50, 2000.
\newblock \doi{10.1145/353323.353382}.
\newblock URL \url{https://www.cs.cornell.edu/fbs/publications/EnfSecPols.pdf}.

\bibitem[Shi et~al.(2026)Shi, He, Wang, Li, Wu, Guo, and Song]{shi2026progent}
Tianneng Shi, Jingxuan He, Zhun Wang, Hongwei Li, Linyu Wu, Wenbo Guo, and Dawn
  Song.
\newblock {Progent}: Securing {AI} agents with privilege control.
\newblock \emph{arXiv}, 2026.
\newblock \doi{10.48550/arXiv.2504.11703}.
\newblock URL \url{https://arxiv.org/abs/2504.11703v3}.
\newblock arXiv v3, 14 May 2026.

\bibitem[Shoham(1993)]{shoham1993agentOrientedProgramming}
Yoav Shoham.
\newblock Agent-oriented programming.
\newblock \emph{Artificial Intelligence}, 60\penalty0 (1):\penalty0 51--92,
  1993.
\newblock \doi{10.1016/0004-3702(93)90034-9}.

\bibitem[{Sigstore}(2026)]{sigstore2026docs}
{Sigstore}.
\newblock Sigstore documentation, 2026.
\newblock URL \url{https://docs.sigstore.dev/}.

\bibitem[{SLSA}(2026)]{slsa2026spec}
{SLSA}.
\newblock Supply-chain levels for software artifacts specification, 2026.
\newblock URL \url{https://slsa.dev/spec/v1.2/}.
\newblock Version 1.2.

\bibitem[Strom and Yemini(1986)]{strom1986typestate}
Robert~E. Strom and Shaula Yemini.
\newblock Typestate: A programming language concept for enhancing software
  reliability.
\newblock \emph{IEEE Transactions on Software Engineering}, SE-12\penalty0
  (1):\penalty0 157--171, 1986.
\newblock \doi{10.1109/TSE.1986.6312929}.
\newblock URL
  \url{https://research.ibm.com/publications/typestate-a-programming-language-concept-for-enhancing-software-reliability}.

\bibitem[{Temporal Technologies}(2026)]{temporalDurableExecution2026}
{Temporal Technologies}.
\newblock What is {Temporal}?, 2026.
\newblock URL \url{https://docs.temporal.io/temporal}.

\bibitem[{The Rust Project}(n.d.)]{rustOwnership2026}
{The Rust Project}.
\newblock What is ownership?
\newblock The Rust Programming Language, n.d.
\newblock URL
  \url{https://doc.rust-lang.org/book/ch04-01-what-is-ownership.html}.

\bibitem[{Treasury Board of Canada
  Secretariat}(2025{\natexlab{a}})]{canada2025admDirective}
{Treasury Board of Canada Secretariat}.
\newblock Directive on automated decision-making, 2025{\natexlab{a}}.
\newblock URL \url{https://www.tbs-sct.canada.ca/pol/doc-eng.aspx?id=32592}.
\newblock Revised 24 June 2025.

\bibitem[{Treasury Board of Canada
  Secretariat}(2025{\natexlab{b}})]{canada2025admScope}
{Treasury Board of Canada Secretariat}.
\newblock Guide on the scope of the directive on automated decision-making,
  2025{\natexlab{b}}.
\newblock URL
  \url{https://www.canada.ca/en/government/system/digital-government/digital-government-innovations/responsible-use-ai/automated-decision-making/guide-scope-directive-automated-decision-making.html}.

\bibitem[Wang et~al.(2024)Wang, Ma, Feng, Zhang, Yang, Zhang, Chen, Tang, Chen,
  Lin, Zhao, Wei, and Wen]{wang2023llmAgentsSurvey}
Lei Wang, Chen Ma, Xueyang Feng, Zeyu Zhang, Hao Yang, Jingsen Zhang, Zhiyuan
  Chen, Jiakai Tang, Xu~Chen, Yankai Lin, Wayne~Xin Zhao, Zhewei Wei, and
  Ji-Rong Wen.
\newblock A survey on large language model based autonomous agents.
\newblock \emph{Frontiers of Computer Science}, 18\penalty0 (6):\penalty0
  186345, 2024.
\newblock \doi{10.1007/s11704-024-40231-1}.
\newblock URL
  \url{https://journal.hep.com.cn/fcs/EN/10.1007/s11704-024-40231-1}.

\bibitem[{World Health Organization}(2024)]{who2025lmmHealthGuidance}
{World Health Organization}.
\newblock Ethics and governance of artificial intelligence for health: Guidance
  on large multi-modal models.
\newblock Technical report, World Health Organization, 2024.
\newblock URL \url{https://www.who.int/publications/i/item/9789240084759}.

\bibitem[Wu et~al.(2024)Wu, Bansal, Zhang, Wu, Li, Zhu, Jiang, Zhang, Zhang,
  Liu, Awadallah, White, Burger, and Wang]{wu2023autogen}
Qingyun Wu, Gagan Bansal, Jieyu Zhang, Yiran Wu, Beibin Li, Erkang Zhu,
  Li~Jiang, Xiaoyun Zhang, Shaokun Zhang, Jiale Liu, Ahmed~Hassan Awadallah,
  Ryen~W. White, Doug Burger, and Chi Wang.
\newblock {AutoGen}: Enabling next-gen {LLM} applications via multi-agent
  conversations.
\newblock In \emph{First Conference on Language Modeling}, 2024.
\newblock URL \url{https://openreview.net/pdf?id=BAakY1hNKS}.

\bibitem[Yao et~al.(2023)Yao, Zhao, Yu, Du, Shafran, Narasimhan, and
  Cao]{yao2022react}
Shunyu Yao, Jeffrey Zhao, Dian Yu, Nan Du, Izhak Shafran, Karthik Narasimhan,
  and Yuan Cao.
\newblock {ReAct}: Synergizing reasoning and acting in language models.
\newblock In \emph{International Conference on Learning Representations}, 2023.
\newblock URL \url{https://arxiv.org/abs/2210.03629v3}.
\newblock ICLR 2023.

\end{thebibliography}
\end{document}